\documentclass[lettersize,journal]{IEEEtran}
\usepackage{easyReview}
\usepackage[normalem]{ulem}
\usepackage{graphicx}
\usepackage{algorithmic}
\usepackage{xcolor}
\usepackage{amsthm}
\usepackage{amssymb}
\usepackage{amsmath,cases}
\usepackage{cite}
\usepackage{color}
\usepackage{subfigure}
\usepackage{subcaption}
\usepackage{booktabs}
\usepackage{balance}
\usepackage{enumitem}
\usepackage{acro}
\usepackage{algorithm}
\usepackage{float}
\usepackage{cite}

\usepackage[utf8]{inputenc}
\DeclareUnicodeCharacter{202F}{\,}
\definecolor{myblue}{RGB}{30,144,255}
\definecolor{amber}{rgb}{1.0, 0.5, 0}
\definecolor{mypink1}{rgb}{0.9, 0.0, 0.7}
\definecolor{mygreen}{RGB}{30,100,50}

\usepackage{color}

\definecolor{darkgreen}{RGB}{0,128,0}

\DeclareAcronym{AO}{
short = AO,
long = absorptive object
}

\DeclareAcronym{RBS}{
	short = RBS,
	long = radio base station
}

\DeclareAcronym{CSI}{
	short = CSI,
	long = channel state information
}

\DeclareAcronym{D2D}{
	short = D2D,
	long = device-to-device
}

\DeclareAcronym{EC}{
	short = EC,
	long = ergodic capacity
}

\DeclareAcronym{FSPL}{
	short = FSPL,
	long = free space path-loss
}

\DeclareAcronym{HAP}{
	short = HAP,
	long = high-altitude platform
}

\DeclareAcronym{HCU}{
	short = HCU,
	long = high-capacity user
}

\DeclareAcronym{HM}{
	short = HM,
	long = Hungarian method
}

\DeclareAcronym{LCUs}{
	short = LCUs,
	long = low-capacity users
}

\DeclareAcronym{LCU}{
	short = LCU,
	long = low-capacity user
}

\DeclareAcronym{LEO}{
	short = LEO,
	long = low-earth orbit
}

\DeclareAcronym{LoS}{
	short = LoS,
	long = line-of-sight
}

\DeclareAcronym{LTE}{
	short = LTE,
	long = long-term evolution
}

\DeclareAcronym{MEC}{
	short = MEC,
	long = mobile edge computing
}

\DeclareAcronym{MC}{
	short = MC,
	long = Multi-connectivity
}

\DeclareAcronym{nLoS}{
	short = nLoS,
	long = non line-of-sight
}

\DeclareAcronym{NOMA}{
	short = NOMA,
	long = non-orthogonal multiple access
}

\DeclareAcronym{NTN}{
	short = NTN,
	long = non-terrestrial network
}

\DeclareAcronym{QoS}{
	short = QoS,
	long = Quality-of-Service
}

\DeclareAcronym{SAGIN}{
	short = SAGIN,
	long = space-air-ground-integrated network
}

\DeclareAcronym{SAT}{
	short = SAT,
	long = satellite
}

\DeclareAcronym{SINR}{
short  = SINR,
long   = signal-to-noise-and-interference ratio
}

\DeclareAcronym{SIC}{
	short  = SIC,
	long   = successive interference cancellation
}

\DeclareAcronym{SC}{
	short  = SC,
	long   = single connectivity
}

\DeclareAcronym{TN}{
	short = TN,
	long = terrestrial network
}

\DeclareAcronym{URLLC}{
short = URLLC,
long = ultra-reliable and low-latency communication
}

\DeclareAcronym{UAV}{
	short = UAV,
	long = uncrewed aerial vehicle
}

\DeclareAcronym{UAVs}{
	short = UAVs,
	long = uncrewed aerial vehicles
}

\begin{document}
	
	\title{
		Resource Allocation and Sharing for UAV-Assisted Integrated TN-NTN with Multi-Connectivity
	}		
	
	\author{
		Abd~Ullah~Khan, 
		Wali~Ullah~Khan, 
		Haejoon~Jung,~\IEEEmembership{Senior Member,~IEEE},
		and 
		Hyundong~Shin,~\IEEEmembership{Fellow,~IEEE}
		
		\thanks{
			A.~U.~Khan, 
			H.~Jung,
			and 
			H.~Shin 
			are with the Department of Electronics and Information Convergence Engineering,
			Kyung Hee University,
			1732 Deogyeong-daero, Giheung-gu,
			Yongin-si, Gyeonggi-do 17104,
			Republic of Korea
			(e-mail: \{abdullah, haejoonjung, hshin\}@khu.ac.kr). A.~U.~Khan is also affiliated with ID Lab, Ghent University, and with National University of Sciences and Technology, Balochistan Campus, Pakistan.
		} 
		\thanks{
			W.~U.~Khan 
			is with the Interdisciplinary Centre for Security, Reliability and Trust (SnT), 
			University of Luxembourg, 
			Luxembourg. (email: waliullah.khan@uni.lu).
		} 
	}

	\maketitle
	\markboth{Khan \textit{\MakeLowercase{et al.}}:
		Resource Allocation and Sharing for UAV-Assisted Integrated TN-NTN with Multi-Connectivity 
	}{ 
		Khan \textit{\MakeLowercase{et al.}}:
		Resource Allocation and Sharing for UAV-Assisted Integrated TN-NTN with Multi-Connectivity 	
	}	
	
	\begin{abstract}
		Unmanned aerial vehicles (UAVs) with multi-connectivity (MC) capabilities efficiently and reliably transfer data between terrestrial networks (TNs) and non-terrestrial networks (NTNs). However, optimally sharing and allocating spectrum and power resources to maintain MC while ensuring reliable connectivity and optimal performance remains challenging in such networks. Channel variations induced by mobility in UAV networks, coupled with the varying quality of service (QoS) demands of heterogeneous devices, resource sharing, and fairness requirements in capacity distribution pose challenges to optimal resource allocation. Thus, this paper investigates resource allocation for QoS-constrained, MC-enabled, dynamic UAVs in an integrated TN-NTN environment with spectrum sharing and fairness considerations. To this end, we consider three types of links: UAV-to-radio base station (RBS), UAV-to-UAV, and UAV-to-HAP. We also assume two types of UAVs with diverse QoS requirements to reflect a practical scenario. Consequently, we propose two algorithms. The first algorithm maximizes the capacity of UAVs-RBS and UAVs-HAP links while ensuring the reliability of the UAV-UAV link. To achieve this, the algorithm maximizes the collective throughput of the UAVs by optimizing the sum capacity of all the UAV-RBS and UAV-HAP links. Next, to provide constant capacity to all links and ensure fairness, we propose another algorithm that maximizes the minimum capacity across all links. We validate the performance of both algorithms through simulation.
	\end{abstract}
	
	


	\section{Introduction}
	\label{sec:1}
	\IEEEPARstart{U}{nmanned} aerial vehicles (UAVs) can sense the environment and transfer application data between terrestrial networks (TNs) and non-terrestrial networks (NTNs) more cost-effectively compared to high-altitude platforms (HAPs) and satellites \cite{TZBYEB:24:IEEE_J_IOT}. Thus, UAVs are expected to play a key role in future integrated terrestrial and non-terrestrial networks (TN-NTN), promoting applications such as autonomous transportation, smart cities, and real-time immersive communication \cite{SQWC:22:IEEE_J_WCOM}. Specifically, the role of UAVs in integrated TN-NTN has become increasingly significant due to their ability to be supported by NTNs via HAPs and TNs through radio base stations (RBS) in cellular and non-cellular networks, given the stringent quality-of-service (QoS) requirements mandated by future applications in terms of data rate and reliability.
	
	Additionally, multi-connectivity (MC)-enabled UAVs can significantly improve the data rate and reliability of integrated TN-NTN. MC is a key technology for 6G networks that has been standardized by the 3rd Generation Partnership Project (3GPP), which is intended to improve reliability and data rates \cite{JML:25:IEEE_J_IOT, ZHZH:25:IEEE_J_TNSE}. UAVs with MC capability can maintain simultaneous connections to multiple network nodes, including ground-based infrastructure and HAPs. This allows them to aggregate resources from various sources and ensure reliability with low latency. Thus, MC can enable numerous novel applications with stringent QoS requirements.
	
	Due to these potentialities, a substantial deployment of UAVs is anticipated in densely populated urban environments in future 6G networks. This large-scale deployment of UAVs will introduce several challenges, including interference, dynamic channel conditions, link reliability, fairness, and satisfaction of QoS requirements \cite{GDMS:22:IEEE_M_COM}. These challenges intensify when considering user heterogeneity and resource sharing, which is a key enabler for realizing integrated TN-NTN 6G networks, as discussed in \cite{JBBJKL:24:IEEE_J_IOT}. This necessitates designing efficient resource sharing and allocation schemes. In particular, resource allocation in a UAV environment under an integrated TN-NTN system is a popular research topic, and various research attempts has been conducted as described below.
	
	
	\subsection{Related literature}
	The authors in \cite{SQWC:22:IEEE_J_WCOM} propose aircraft-assisted integrated TN-{NTN}. The authors in \cite{ZYM:23:IEEE_J_COML} discuss the communication links between UAVs and HAPs, and the associated advantages and challenges. The authors further present the network architectures for ad-hoc, cell-free, and integrated access and backhaul, considering UAVs as the key players. In \cite{JJMY:22:IEEE_J_VT}, packet forwarding between ground stations through UAVs is investigated using radio-frequency or free-space optical links. The authors in \cite{GDMS:22:IEEE_M_COM} present the opportunities and challenges in UAV-assisted integrated TN-{NTN} and elaborate on some interesting use cases. They further review the relevant standardization activities and present unexplored research problems. Additionally, they consider how TN and NTN can be efficiently manipulated to provide services to each other. Further, various scenarios for UAVs in NTNs are discussed in \cite{CSA:22:IEEE_M_IOT}, covering NTN architecture and the challenges associated with connectivity and QoS requirements. The authors also present some open research problems. The authors in \cite{MMB:24:IEEE_M_IOT} comprehensively survey the UAV-driven cluster-based data-gathering techniques for heterogeneous robotics and wireless sensor networks. The authors in \cite{TZBYEB:24:IEEE_J_IOT} analyze the network structure of NTNs from the perspective of the significance of timeliness in various applications. The authors in \cite{GASF:21:IEEE_M_NET} delve into the constraints offered to 5G new radio (NR) by NTN features. In this regard, the limitations and advantages of the NR technology in realizing 6G services are explored, and open problems are provided. The study in \cite{UAMW:19:IEEE_M_WCOM} overviews the key challenges and design techniques of commercial use of UAVs and discusses the role of cellular networks to support UAVs leveraging network intelligence and new tools from machine learning. In \cite{MGRD:22:IEEE_C_GC}, the authors assess integrated TN-{NTN} taking into account cellular network jointly with HAPs to provide link connection to ground users and UAVs.
	
	Aside from this, the authors in \cite{LMXSSQ:21:IEEE_J_VT}, \cite{J:23:IEEE/ACM_J_NET}, and \cite{DJJJ:25:IEEE_J_MASS} use data-driven approach in resource allocation. On the other hand, the continuous coverage problem is considered in \cite{LMXSSQ:21:IEEE_J_VT} for high dynamic multi-UAV communication networks. Also, the imperfect channel state information (CSI) caused by the high mobility is overcome by proposing a deep reinforcement learning method, which jointly optimizes channel and power allocation. The authors in \cite{J:23:IEEE/ACM_J_NET} address a mobility-resilient spectrum sharing framework for swarm UAVs, addressing the challenge of interference with incumbent systems that use directional antennas. They utilize a data-driven, three-phase approach to manage UAV transmission power, flight, and user association. The authors in \cite{DJJJ:25:IEEE_J_MASS} propose a spectrum sharing scheme based on multi-agent deep reinforcement learning (DRL) to address spectrum scarcity in scenarios where UAV networks coexist with terrestrial networks. 
	
	In \cite{KYSC:23:IEEE_J_CCN} and \cite{KHYA:21:IEEE_J_CCN}, cognitive radio-based resource allocation has been investigated in many prior arts. Specifically, the authors in \cite{KYSC:23:IEEE_J_CCN} consider channel assignment and power allocation in non-orthogonal multiple access (NOMA)-enabled cognitive satellite-UAV-terrestrial networks operating with imperfect CSI. Similarly, the authors in \cite{KHYA:21:IEEE_J_CCN} propose resource allocation for a cognitive radio-UAV system by jointly optimizing UAV route/position, power, and time allocation under interference constraints and a probabilistic channel. Moreover,  spectrum sharing and resource allocation for a UAV network are investigated in \cite{MMNRHE:24:IEEE_J_COM} and \cite{WQLA:22:IEEE_J_TV}.
	The authors in \cite{MMNRHE:24:IEEE_J_COM}  introduce an enhanced licensed shared access framework designed to address spectrum scarcity in 6G networks by integrating UAV-based spectrum sensing and a dynamic spectrum auction. 
	On the other hand, the authors in \cite{WQLA:22:IEEE_J_TV} maximize energy efficiency in UAV-assisted vehicular networks with spectrum sharing between UAV-to-vehicle (U2V) and vehicle-to-vehicle (V2V) links, which introduces interference and burdens infrastructure. 

	UAV-aided short-packet NOMA systems are studied under realistic conditions of imperfect CSI and imperfect successive interference cancellation (SIC) in \cite{H:22:IEEE_J_CCN}. To combat performance degradation, they formulate and solve optimization problems for power allocation, joint power allocation, and coding rate maximization of throughput. Additionally, \cite{ZCKM:19:IEEE_J_WL} and \cite{MNYLX:20:IEEE_J_VT} use mobile edge computing (MEC) in resource allocation to a UAV network and UAV-assisted communication. The authors \cite{ZCKM:19:IEEE_J_WL} consider the complex sum power minimization problem in multi-UAV MEC networks by jointly optimizing user association, power control, computation capacity, and UAV location. Whereas, the authors in \cite{MNYLX:20:IEEE_J_VT} maximize energy efficiency in UAV-assisted MEC by jointly optimizing UAV trajectory and resource allocation such as user transmit power and computation load.

	Energy harvesting-assisted resource allocation is also an important aspect in UAV networks, as noted in \cite{YZCC:21:IEEE_J_IOT}, \cite{YDKLR:19:IEEE_J_COM}, and \cite{MDTD:18:IEEE_J_WCL}. The authors in  \cite{YZCC:21:IEEE_J_IOT} consider resource allocation in energy harvesting-based device-to-device (D2D) communication underlying UAV-assisted networks, focusing on maximizing energy efficiency. 
	In \cite{YDKLR:19:IEEE_J_COM}, the authors investigate the joint design of three-dimensional aerial trajectory and wireless resource allocation for solar-powered UAV communication systems to maximize sum throughput over time.
	Moreover, the authors in  \cite{MDTD:18:IEEE_J_WCL}  propose a real-time, low-complexity algorithm to maximize energy efficiency in a UAV-assisted D2D wireless information and power transfer system by jointly optimizing energy harvesting time and power control.

	Ensuring secure communications by optimizing resource allocation in UAV communications is also of paramount importance, as discussed in \cite{XZRKJ:20:IEEE_J_COM} and \cite{RZLKYNYA:20:IEEE_J_COM}. 	The authors in \cite{XZRKJ:20:IEEE_J_COM} maximize energy efficiency in a secure UAV communication system with an information UAV and a multi-antenna jammer UAV, jointly optimizing the information UAV's trajectory, resource allocation, and the jammer's policy. Furthermore, the authors in \cite{RZLKYNYA:20:IEEE_J_COM} tackle the complex, non-convex problem of secure resource allocation and trajectory design for multi-UAV systems with multiple eavesdroppers to maximize the average minimum secrecy rate per user.
	\renewcommand{\arraystretch}{2}
	\begin{table*}[t]
		\centering
		\caption{\textcolor{blue}{Comparison with existing literature on MC}}
		\label{tab_lit_comparison}
		\begin{tabular}{|p{1cm}|p{1.5cm}|p{2cm}|p{2cm}|p{1.5cm}|p{1cm}|p{4.5cm}|} \hline
			\textbf{Ref}.                                                         & \textbf{MC}  & \textbf{UAVs' heterogeneity} & \textbf{Differentiated QoS}  & \textbf{Spectrum sharing} & \textbf{Fairness} & \textbf{Optimization Target}                                                  \\ \hline
			\cite{Wu2024diffusionMetaverseUAVdc   }     & Yes & No                  & No                           & No                          & No                   & To minimize frame latency   and energy                                  \\ \hline
			\cite{Hoang2023DRLMECdualConnectivity   }   & Yes & No                  & No                             & No                          & No                   & To minimize  power complexity \\ \hline
			\cite{Li2019efficientResourceUAVBSdc   }    & Yes & Yes                 & No                             & No                          & No                   & Throughput enhancement                                             \\ \hline
			\cite{Sae2020dualMNOuavReliability   }      & Yes & Yes                   & No                               & No                          & No                   & To improve links reliability                               \\ \hline
			\cite{Khalili2022UAVRISdualConnectivity   } & Yes & Yes                 & No                                 & No                          & No                   & To minimize transmit   power                                      \\ \hline
			\cite{Cheng2024learningMCUAV   }            & Yes & No                   & No                               & No                          & No                   & To maximize long-term   network utility                              \\ \hline
			\cite{Zhang2023uavNOMAdualConnectivity   }  & Yes & No                 & No                               & No                          & No                   & To maximize sum   rate                                      \\ \hline
			\cite{Cao2024outageMCUAVmmWave   }          & Yes & No                     & No                               & No                          & No                   &    To reduce outage probability                                                                  \\ \hline
			\cite{Amorim2019dualNetworkC2   }           & Yes & Yes                 & No                               & No                          & No                   & To improve links   reliability \\ \hline
			\cite{Lai2024fuzzyHOMCcellularUAV   }       & Yes & No             & No                               & No                          & No                   & To improve link quality                                              \\ \hline
			\cite{Sadovaya2024MCoffloadingNTN   }       & Yes & No   & No                               & No                          & No                   & To improve service continuity                                     \\ \hline
			\cite{RamirezArroyo2025MCrural5Gsat   }     & Yes & No              & No                               & No                          & No                   &   To maximize capacity   \\ \hline   
			\textbf{This work}                                                    & \textbf{Yes} & \textbf{Yes}              & \textbf{Yes}                              & \textbf{Yes}                         & \textbf{Yes}                  & \textbf{To maximize the minimum and maximum capacities}                     \\ \hline                                                            
		\end{tabular}
	\end{table*}

	In addition, various advanced designs and optimization techniques have been proposed for resource allocation, interference management, and system design for UAV networks, as in  \cite{JYA:19:IEEE_J_WCOM,ZWWH:21:IEEE_J_JSAC, D:22:IEEE_M_COMS, SVGK:20:IEEE_J_COM}. To be specific, the authors in \cite{JYA:19:IEEE_J_WCOM} investigate dynamic resource allocation (e.g., user, power level, subchannel selection) for multiple UAVs acting as aerial base stations, operating independently without information exchange. The work in \cite{ZWWH:21:IEEE_J_JSAC} addresses resource allocation for UAV-based emergency wireless communications in disaster areas with rapidly changing channels by proposing a novel heterogeneous Fisher-Snedecor composite fading channel model and adaptive power and bandwidth allocation schemes. Also, the work in  \cite{D:22:IEEE_M_COMS} addresses the 3GPP Release 15 study, which investigated enhanced long-term evolution (LTE) support for connected UAVs. The study identified significant challenges, such as increased uplink and downlink interference due to the high likelihood of line-of-sight propagation of aerial vehicles to multiple neighboring cells. This work resulted in the standardization of Release 15 features designed to address several issues within the LTE network. The authors in \cite{SVGK:20:IEEE_J_COM} propose a new composite channel model for UAV-to-ground communications that accounts for both scattering and shadowing, addressing scenarios where line-of-sight conditions are not met. Additionally, they introduce a low-complexity UAV selection policy that utilizes slower-varying shadowing information rather than instantaneous channel state information.
	
	\textcolor{blue}{\subsubsection{MC-enabled UAV communication}	
		Though substantial literature exists on UAV resource allocation as described above, MC-enabled UAV communication remains an emerging research that has been investigated in only a few studies. For instance, \cite{Wu2024diffusionMetaverseUAVdc}  develops a UAV-enabled MEC architecture with MC using a diffusion-model-based offloading algorithm that jointly optimizes offloading and communication/compute resources to minimize frame latency and energy. Similarly, \cite{Hoang2023DRLMECdualConnectivity } designs a Lyapunov-guided DRL framework to jointly optimize MC offloading and computing decisions in a UAV-assisted MEC system to minimize long-term system power. 
		\cite{Zhang2023uavNOMAdualConnectivity } Considers a UAV-assisted uplink NOMA system with MC and jointly optimizes UAV placement and transmit power using a hybrid fractional programming to maximize weighted sum rate. Similarly, \cite{Cao2024outageMCUAVmmWave } provides analytical outage probability expressions for UAV-assisted MC mmWave links in urban environments, accounting for blockage and spatial diversity.
		\cite{Amorim2019dualNetworkC2 } proposes to use MC to two independent networks to improve the links reliability of UAVs.
		\cite{Lai2024fuzzyHOMCcellularUAV } develops a fuzzy-logic handover decision scheme for cellular-connected UAVs under MC, aiming to improve link quality during mobility.	\cite{Sadovaya2024MCoffloadingNTN } investigates how MC and traffic offloading between NTN platforms (e.g., HAP/UAV) and terrestrial nodes can improve service continuity.
		\cite{RamirezArroyo2025MCrural5Gsat } demonstrates and analyzes MC between terrestrial networks and LEO satellite links for smart agriculture scenarios.
		\cite{Li2019efficientResourceUAVBSdc } proposes a genetic algorithm-based scheme to efficiently allocate radio resources to UAV-BS-assisted heterogeneous network with MC. Similarly, \cite{Khalili2022UAVRISdualConnectivity } studies RIS-assisted heterogeneous network with UAV relays and MC, and uses multi-agent deep Q-network to jointly control UAV trajectory, RIS configuration, and resource allocation.
		\cite{Sae2020dualMNOuavReliability } conducts extensive field measurements of UAV connectivity using two mobile network operators and evaluates reliability benefits of MC.
		\cite{Cheng2024learningMCUAV } formulates a joint UAV–user association and power/channel allocation problem in a MC-enabled UAV network and solves it using a multi-agent hybrid DRL.}

	Despite the extensive literature on allocating resources to UAV-assisted networks, several emerging challenges in the context of integrated TN-NTN remain unexplored. Specifically, the existing literature generally disregards device heterogeneity, which is central to UAV-assisted integrated TN-NTN. Similarly, as shown in Table \ref{tab_lit_comparison}, existing literature that considers MC does not address the joint optimization of spectrum allocation and power control within a UAV-assisted integrated TN–NTN MC framework. Furthermore, when considering heterogeneity alongside MC, device mobility, dynamic resource sharing, stringent QoS requirements, and fairness, a complex resource allocation problem arises that, to our knowledge, has not yet been investigated. Thus, developing an optimal framework for sharing and allocating resources that addresses these interconnected challenges is a critical yet unexplored area in UAV-assisted integrated TN-NTN.


	To fill this research gap, we propose a novel resource sharing and allocation framework wherein we consider differentiated QoS-constrained MC-enabled heterogeneous UAVs in integrated TN-{NTN}. To this end, we consider three types of UAV links, such as UAV-RBS, UAV-HAP, and UAV-UAV. The UAV-RBS link is supported by major TN-NTN links, while the UAV-HAP links operate through major NTN-NTN links. Lastly, the UAV-UAV links are enabled by the local NTN-NTN links. Furthermore, we consider differentiated QoS requirements for these links, implying that high capacity is required for major links (i.e., UAV-RBS and UAV-HAP), as they have to transfer a bulk of data. On the other hand, high reliability is required for local (i.e., UAV-UAV) links, as they have to transfer safety-critical information. Moreover, we consider dynamic spectrum sharing among various links in a dynamic environment and circumvent the challenge caused by rapidly changing channels by performing resource sharing and allocation based on slow fading conditions and statistics of the channel rather than relying on instantaneous CSI. Accordingly, we formulate our optimization problem considering both the sum and minimum capacity for maximization.
	
	By jointly considering the heterogeneity of UAVs with MC, and exploiting only the large-scale fading statistics, while accounting for small-scale fading and fairness, this work offers a novel resource sharing and allocation framework for QoS-constrained heterogeneous MC-enabled UAVs in integrated TN-{NTN}. Table \ref{tab_lit_comparison} presents a comparative analysis between the proposed work and existing literature. The contributions are summarized below.
	\begin{itemize}
		\item The resource sharing and allocation process is formulated as a novel optimization problem that reflects the unique characteristics of UAV communications, including time-varying channels and heterogeneity with differentiated QoS requirements. Additionally, a novel Hungarian method (HM)-based two-step optimization approach is proposed that determines the optimal spectrum sharing pattern.
		\item Two algorithms are proposed to tackle the complex optimization problem. The first algorithm maximizes the capacity of the major links, i.e., UAV-RBS and UAV-HAP, while making the local links, i.e., UAV-UAV, guaranteed reliable. The second algorithm ensures constant capacity for all major links and enhances the minimum capacity, thereby promoting fairness. Both algorithms are extensively analyzed under various scenarios and compared against the state-of-the-art.

	\end{itemize}
	The remaining paper is organized as follows. The next section presents the system model. Section \ref{sec_Sum_cap_max} presents the sum capacity maximization, while Section \ref{sec_min_cap_max} presents the minimum capacity maximization problems. Section \ref{sec_results} presents simulation results, and Section \ref{sec_conc} concludes the paper. 
	Table \ref{tab_list_of_notations} presents the list of notations and their definitions.
	
	\renewcommand{\arraystretch}{1.25}
	\begin{table}[t]
		\centering
		\caption{List of key notations and abbreviations}
		\label{tab_list_of_notations}
		\begin{tabular}{|p{0.9cm}|p{6.8cm}|}
			\hline
			
			\hline
			\textbf{Notation} & \textbf{Definition} \\
			\hline
			
			$\alpha_{x,y}$  & Link between $x$ and $y$ where $x,y \in \{i,j,R,H\}$ \\\hline
			$C_i^h$ &  capacity of the $i$th HCU\\ \hline
			
			$C_{i,j}^h$ 	& Capacity of the $i$th HCU when sharing spectrum with the $j$th LCU \\ \hline
			
			$C^*_{i,j}$ & Maximum capacity of $i$th HCU when sharing spectrum with the $j$th LCU	 \\ \hline
			$C^h_0$   & Minimum capacity requirement of HCUs \\ \hline	
			
			$\gamma_{i,H}^h$ & The SINR obtained at HAP for the $i$th HCU
			\\ \hline
			$\gamma_{i,R}^h$  &  The SINR obtained at RBS for the $i$th HCU \\\hline
			$\gamma_{j,j}^\ell$  &  The SINR obtained at the $j$th LCU in LCU-LCU pair\\\hline
			$\gamma_0^\ell$ & Minimum SINR for LCUs for reliable link establishment\\ \hline	
			
			HCU & High-capacity user \\\hline	
			HM & Hungarian method \\ \hline
			$h_{i,R}$ & Channel power gain between the $i$th HCU and RBS \\\hline
			$h_{j,R}$  & The interference channel from the $j$th LCU to RBS \\\hline
			$h_{i,j}$  & The interference channel from the $i$th HCU to $j$th LCU \\\hline	$h_{j,j}$  & The channel between the $j$th LCU-LCU pair \\\hline
			$I$ &  The total number of HCUs\\\hline
			$J$ &  The total number of LCUs\\\hline
			LCU & Low-capacity user \\\hline
			$L_p$ & Path-loss factor \\\hline
			
			MC & Multi-connectivity \\\hline
			
			$P_i^h$  &  Transmitting power of the $i$th HCU \\\hline
			
			$P^\ell_j$  & Transmitting power of the $j$th LCU \\\hline
			
			$P_{m}^h$ 		& Maximum transmitting power of HCU \\ \hline	
			
			$P_{m}^\ell$  		& Maximum transmitting power of LCU \\ \hline	
			
			$P_i^{h^*}$  & Optimal transmitting power for HCU \\ \hline

			$P_j^{\ell^*}$  &  Optimal transmitting power for LCU \\ \hline	
			
			$P_o$ & The acceptable outage probability \\ \hline	
			
			SC & Single-connectivity \\\hline
			
			$\mu_{i,j}$ 	& Spectrum allocation index \\ \hline	
			
			$\chi_{x,y}$  & The log-normal shadow fading random variable between $x,y \in \{i,j,R,H\}$\\\hline
			
			\hline
		\end{tabular}
	\end{table}
	
	\begin{figure}[t!]	
		\centering
		\includegraphics[width=0.37\textwidth]{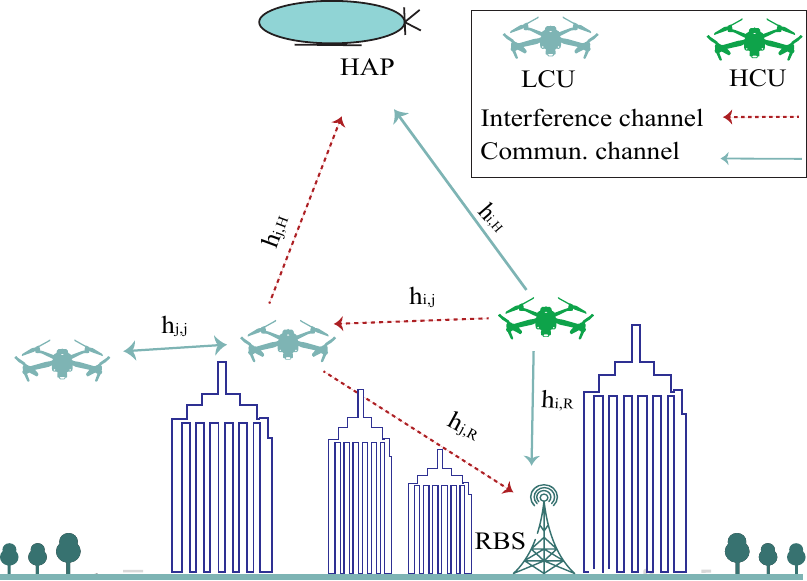}
		\caption{
			System model of the UAV-assisted integrated TN-{NTN}.
		}
		\label{fig:system_model}
	\end{figure}

	
	\section{System Model}
	We consider a UAV-assisted integrated TN-{NTN} architecture in a dense urban environment, as shown in Fig. \ref{fig:system_model} based on \cite{GDMS:22:IEEE_M_COM, WZYQ:25:Elsevier_ENGG}. There are $I$ UAVs that need high-capacity UAV-RBS and UAV-HAP links. These $I$ UAVs are called high-capacity users (HCUs). On the other hand, there are also $J$ pairs of UAVs performing local UAV-UAV communications. They are called low-capacity users (LCUs), as in \cite{GDMS:22:IEEE_M_COM, QJA:25:wiley:TETT}. In order to enhance capacity, HCUs utilize both UAV-RBS and UAV-HAP links through MC. All UAVs can perform UAV-RBS, UAV-HAP, and UAV-UAV communications \cite{R:24:Elsevier_CN}. Hence, HCUs and LCUs may represent the same UAVs in this paper.
	
	Suppose that the set of HCUs is denoted by $ \mathcal{I} = \{1,\dots, I\}$, whereas the set of LCUs pairs is denoted by $\mathcal{J} = \{1,\dots, J\}$. To enhance spectral efficiency, the spectrum allocated orthogonally to HCUs is also utilized by LCUs. As in \cite{BEI:22:IEEE_J_AES}, the channel between the $i$th HCU and RBS is given by 
	\textcolor{blue}{	\begin{equation}\label{eq_1}
			h_{i,R}(t) = L_p \hspace{0.5mm}D_{i,R}(t)^{-\phi} \hspace{0.5mm}\chi_{i,R}\hspace{0.5mm} g_{i,R}^{(s)}e^{j2\pi (f_{d,iR})t},
	\end{equation}}
	where $L_p$ represents the path-loss factor, which is a constant quantity, while $D_{i,R}$ indicates the distance between the $i$th HCU and the RBS, and $\phi$ is the path loss exponent.
	 \textcolor{red}{Further, $g_{i,R}^{(s)}$ is the random small-scale fading power gain, representing the envelope of multipath-induced fading arising from propagation delays and Doppler shifts \cite{F:22:B_JWS}. Specifically, $g^{(s)}_{i,R}$ is assumed to follow exponential distribution with the mean of one. }
	Similarly, $\chi_{i,R}$ denotes the random shadowing gain, which is characterized by a log-normal distribution with a standard deviation $\xi$. \textcolor{blue}{$f_{d,iR}$ is the Doppler shift in the link given by $f_{d,iR} = \frac{v}{\lambda}\cos\theta_{i,R}$ with $\lambda$ being the carrier wavelength and $\theta_{i,R}$ the angle between the UAV velocity $v$ and the UAV–BS line-of-sight (LoS) direction. Since the Doppler effects appear only as the phase factor $e^{j2\pi (f_{d,iR})t}$, under the standard assumption of proper carrier-frequency/Doppler compensation with no significant inter-carrier interference, and using statistical channel parameters instead of instantaneous CSI for fast fading, we have $|h_{i,R}(t)| = |h_{i,R}|$ \cite{liu2020angular}.}
	In addition, $h_{j,j}$ is the channel between the $j$th UAV-UAV pair, and $h_{i,H}$ is the channel between the $i$th HCU and the HAP. Moreover, $h_{j,R}$ is defined to be the interference channel between the $j$th LCU and RBS, where the interference originates from the $j$th LCU to the RBS. Also, $h_{i,j}$ is the interference channel from the $i$th HCU to the $j$th LCU, and $h_{j,H}$ is the interference channel from the $j$th LCU to the HAP. The respective $\chi_{x,y}$, $D_{x,y}^{-\phi}$, and $g_{x,y}^{(s)}$ are also defined in the same way, where $x,y \in \{i,j,R,H\}$.

\textcolor{red}{It is worth noting that the small-scale fading component $g^{(s)}_{i,R}$ models the envelope variation arising from multipath propagation. The received signal at RBS consists of multiple signal components traveling along different paths. The signal corresponding to the $l$-th path has a propagation delay $\tau_l$, a Doppler shift $f_{d,l}$, an amplitude $a_l$, and a random phase $\theta_l$. The superposition of these components at the receiver can be expressed as
	\begin{equation}
		h(t) = \sum_{l=1}^{L} a_l e^{j\left(\theta_l + 2\pi f_{d,l} t\right)}
		\delta(\tau - \tau_l),
		\label{eq:multipath_model}
	\end{equation}
	where $L$ denotes the number of multipath components.
	The envelope $|h(t)|$ of this superposition follows Rayleigh statistics, which justifies modeling its magnitude using an exponential distribution, corresponding to Rayleigh fading \cite{F:22:B_JWS}.
Therefore, modeling the statistical envelope $g^{(s)}_{i,R}$, rather than explicitly tracking individual multipath parameters, is not only	mathematically rigorous and computationally tractable but also a standard practice in mobile wireless systems for resource allocation problems~\cite{F:22:B_JWS,9593154,sreekumar2015distributed}.}

Aside form this, it is assumed that RBS can estimate the path loss and shadowing values for HCU-RBS and LCU-RBS links. To measure the channel parameters for the LCU-LCU and HCU-LCU links, LCU receivers are used to measure the channel parameters for the links between LCU-LCU and HCU-LCU, and the measured values are regularly shared with the RBS. UAVs are mobile, which causes the channel to experience fast fading. \textcolor{red}{ The rapid fluctuations in fast fading make it challenging for the RBS to compute its exact values. However, RBS can calculate the statistical parameters of the fast fading channels.}
With the above assumptions, the signal-to-noise-and-interference ratio (SINR) obtained at the RBS for the $i$th HCU is given by \cite{MA:11:IEEE_J_COM}
\begin{equation} \label{eq_sinr_i_R}
	\gamma_{i,R}^h = \frac{P_i^h \ |h_{i,R}|}{P_N + \sum\limits_{j \in J}\ \mu_{i,j}\ P_{j}^\ell\ |h_{j,R}|},
\end{equation}
where $P_i^h$ is the transmitting power of the $i$th HCU, $P^\ell_j$ represents the transmitting power of the $j$th LCU, and $P_N$ corresponds to the noise power. Additionally, $\mu_{i,j}$ is the spectrum allocation index, where $\mu_{i,j}=1$ means that the $j$th LCU uses the spectrum originally allocated to the $i$th HCU. Contrarily, $\mu_{i,j}=0$ indicates that no LCU uses the spectrum allocated to HCU. In the former case, the SINR at the $j$th LCU in the corresponding LCU-LCU pair is given by \cite{MA:11:IEEE_J_COM}
\begin{equation}\label{eq_sinr_j_j}
	\gamma_{j,j}^\ell= \frac{P_j^\ell \ |h_{j,j}|}{P_N + \sum\limits_{i \in I} \mu_{i,j} \ P_i^h \ |h_{i,j}|}.
\end{equation} 
It is assumed that HCUs use the same transmission power, $P_i^h$, when communicating with HAP. Hence, the SINR obtained at HAP for the $i$th HCU is given by 
\begin{equation} \label{eq_sinr_i_H}
	\gamma_{i,H}^h = \frac{P_i^h \ |h_{i,H}|}{P_N + \sum\limits_{j \in J} \ \mu_{i,j} \ P_{j}^\ell \ |h_{j,H}|}.
\end{equation}

The expression for the capacity of the $i$th HCU at RBS and HAP can thus be respectively given as \cite{WJXGC:23:IEEE_J_COML}
\begin{equation} \label{eq_erg_cap_1}
	C_{i,R}^h = \mathbb{E} \left[ \log_2 \left( 1 + \gamma_{i,R}^h \right) \right],
\end{equation}
\begin{equation}\label{eq_erg_cap_2}
	C_{i,H}^h =	\mathbb{E} \left[ \log_2 \left( 1 + \gamma_{i,H}^h \right) \right],
\end{equation}
where the notation $\mathbb{E}[.]$ is the statistical expectation. 
Writing in combined form, we have
\begin{equation}
	C_i^h =	 	C_{i,R}^h + C_{i,H}^h.
\end{equation}


\textcolor{blue}{\textit{UAV Energy Consumption}: Each $x_{th}$ UAV where $x \in \{I,J\}$ consumes energy due to both propulsion and communication. 
	Let $P^{\mathrm{prop}}_x(v_x)$ denotes the propulsion power required to sustain the flight speed $v_x$, and let $P^{\mathrm{comm}}_x$ denotes the communication power. 
	To simplify the propulsion-energy modeling, the position and velocity of each UAV are considered constant within a time slot, which allows $P^{\mathrm{prop}}_i$ to be treated as a constant during that interval. 	
	Accordingly, the total power consumption of UAV $x$ at slot $t$ is given by
	\begin{equation}
		P^{\mathrm{tot}}_x(t) = P^{\mathrm{prop}}_x\!\left(v_x(t)\right) + P^{\mathrm{comm}}_x(t),
		\label{eq:UAV_total_power}
	\end{equation}
	and the corresponding per-slot energy expenditure is
	\begin{equation}
		E_x(t) = P^{\mathrm{tot}}_x(t)\,\tau ,
		\label{eq:UAV_energy_slot}
	\end{equation}
	where $\tau$ denotes the slot duration.	 
	Consequently, over a time horizon of $T$ slots, the accumulated energy consumption of UAV $i$ is
	\begin{equation}
		E_i^{\mathrm{tot}}
		= \sum_{t=1}^{T} E_i(t)
		= \tau \sum_{t=1}^{T} 
		P^{\mathrm{tot}}_i(t).
		\label{eq:UAV_total_energy}
\end{equation}}

\section{Sum HCU Capacity Maximization} \label{sec_Sum_cap_max}
In this section, we develop an efficient resource sharing and allocation framework that allocates spectrum and power to UAVs to improve their communications performance under the integrated TN-{NTN} characteristics. Our framework relies only on the large-scale channel parameters. 
\textcolor{red}{The reason is that large-scale CSI varies slowly even at higher velocities and frequencies, and can be reliably estimated from locations or long-term measurements, as it remains stable over a long time duration, making it suitable for centralized optimization \cite{Goldsmith2005Wireless,8787874}. 
	Furthermore, in the ergodic capacity calculation in \eqref{eq_erg_cap_1} and \eqref{eq_erg_cap_2}, instantaneous small-scale fading is averaged out, and only statistical knowledge of fast fading is required. Therefore, large-scale CSI is sufficient for accurate performance modeling \cite{dulek2014power}. 
	In fact, large-scale fading and LoS/NLoS conditions are determined by link distance and environmental shadowing, which evolve only after UAV displacements of tens of meters. 
	This corresponds to update intervals on the order of several hundred milliseconds for practical UAV speeds \cite{8787874}. \\
	Hence, through large-scale CSI, robust and slowly-updated decisions can be achieved.
	  }

Furthermore, we consider diverse QoS requirements for major and local links, implying that UAV-RBS and UAV-HAP links require high capacity, while UAV-UAV connections require high reliability. 
\textcolor{red}{\textit{Major and local links are characterized by their roles and traffic characteristics. Major links correspond to the connections between HCUs and the infrastructure nodes, i.e., RBS and HAP. 
		As a result, major links support high capacity communications, e.g., high spectral efficiency and large average throughput, to efficiently deliver mission data, sensing information, and user payloads to the TN and NTN segments.
		From the functionality point of view, major links can be thought of as backhaul/access links. \\
		In contrast, local links are mainly used for control and coordination. From the functionality point of view, local links can be thought of as links meant for exchanging trajectory updates, collision-avoidance information, cluster-head signaling, and short cooperative forwarding messages. 
		The volume of information on these links is limited; however, their reliability is critical, and their failure causes significant network performance degradation, e.g., loss of cluster coordination, safety degradation, interruption of relaying, etc. 
		Therefore, in the literature, local links are characterized by reliability and connectivity requirements, rather than by very high data-rate demands \cite{lim2024uav}}. 
}

Subsequently, we optimize the sum capacity of $I$ HCUs under the constraint of reliability for every LCU. Furthermore, we fix a minimum requirement threshold in terms of capacity for each HCU so that each HCU can achieve a minimum QoS. 
Therefore, the resource allocation problem in the integrated TN-NTN is formulated as
\begin{subequations} \label{eq_opt_1}
	\begin{alignat}{2}
		\max_{\substack{\{\mu_i,j\}  \{P^{h}_{i}\}, \{P^{\ell}_{j}\}
		}}{} \quad & \sum_{i\in I} C_i^h, \\
		\text{s.t.} \quad & C_i^h \geq C^h_0, \forall i \in I,  \label{eq_opt_1b} \\
		& \text{Pr}\left\{\gamma_{j,j}^\ell \leq \gamma_0^\ell  \right\} \leq P_o, \forall j \in J, \label{eq_opt_1c} \\
		& 0 \leq P_i^h \leq P_{m}^h,  \forall i \in I,  \label{eq_opt_1d} \\
		& 0 \leq P_j^\ell \leq P_{m}^\ell, \forall j  \in J, \label{eq_opt_1e} \\
		& \sum_{i\in I} \mu_{i,j} \leq 1, \mu_{i,j} \in \left\{ 0, 1 \right\} \forall j \in J, \label{eq_opt_1f} \\
		& \sum_{j \in J} \mu_{i,j} \leq 1, \mu_{i,j} \in \left\{ 0, 1 \right\} \forall i \in I, \label{eq_opt_1g}\\
		& 		\textcolor{blue}{E_x^{\mathrm{tot}} \le E_x^{\max} \ \ \forall x \in \{I, J\} \label{eq_opt_1h}}.
	\end{alignat}
\end{subequations}
where $C^h_0$ represents the minimum required capacity for HCUs, while $\gamma_0^\ell$ is the minimum SINR at which LCUs can establish reliable connections.
$P_o$ signifies the tolerable outage probability between LCU-LCU connections. 
$P_{m}^h$ shows the maximum transmission power of HCU,  
$P_{m}^\ell$ denotes the maximum transmission power of LCU, and $E_x^{\max}$ denotes the available battery capacity of UAV.

Moreover, constraint \eqref{eq_opt_1b} represents the minimum capacity requirements for each HCU, and \eqref{eq_opt_1c} corresponds to the minimum reliability requirements for each LCU. Similarly, \eqref{eq_opt_1d} and \eqref{eq_opt_1e} introduce limitations on the transmission power of HCUs and LCUs, ensuring that their transmit powers do not exceed the maximum limit. Constraints \eqref{eq_opt_1f} and \eqref{eq_opt_1g} formalize the one-to-one pairing assumption between HCUs and LCUs in spectrum sharing, implying that a single HCU's spectrum can be shared with only one LCU, and that an LCU can access the spectrum of just one HCU.
This assumption simplifies complex interference scenarios in UAV-assisted integrated TN-NTN and provides a foundation for analyzing complex resource-sharing and allocation challenges in such scenarios.

Our optimization approach offers an innovative formulation that considers the unique features of mobility-induced time-varying multipath channels -- characterized by Doppler shifts and rapidly fluctuating small-scale fading -- as well as the different QoS requirements for major and local links. However, the optimization presents a significant challenge due to its highly nonconvex nature, which is attributable to the combinatorial nature of the problem and the complex structure of the objective function.
We propose a two-step solution to tackle this optimization problem. Our initial step exploits the inherent decoupling of power allocation and spectrum sharing pattern design. Since constraints \eqref{eq_opt_1f} and \eqref{eq_opt_1g} ensure a one-to-one pairing between HCUs and LCUs, interference can be isolated within each pair. This enables us to analyze each pair individually. For each HCU-LCU pair, we determine the optimal power allocation that maximizes the HCU's capacity while ensuring the LCU's reliability requirements are met.  Second, we evaluate the feasibility of spectrum sharing and power allocation for each HCU-LCU pair. This step involves checking whether the achieved capacity for the HCU fulfills the minimum capacity requirements specified in \eqref{eq_opt_1b}. Any pair that does not fulfill these requirements is deemed infeasible and excluded from consideration for spectrum sharing and power allocation pairing.




\subsection{Power Allocation to Single HCU-LCU Pair}
This sub-section focuses on determining the optimal power allocation for every feasible LCU and HCU shared-spectrum pair, leveraging the one-to-one pairing assumption introduced earlier. For a given spectrum-sharing pattern in which the $j$th LCU shares the band with the $i$th HCU, the power-allocation problem for this HCU-LCU pair is simplified as \begin{subequations}
	\begin{alignat}{2}\label{eq_opt_2}
		\max_{P^{h}_{i}, P^{\ell}_{j}} \quad &  C_i^h \\
		\text{s.t.} \quad 
		& \text{Pr}\left\{\gamma_{j,j}^\ell \leq \gamma_0^\ell  \right\} \leq P_o, \label{eq_opt_2a}\\
		& 0 \leq P_i^h \leq P_{m}^h, \\
		& 0 \leq P_j^\ell \leq P_{m}^\ell.
	\end{alignat}
\end{subequations}
Given the power allocation problem for a single pair formulated above, the reliability constraint for the jth LCU can be expressed as follows (see Appendix A for proof):
\begin{equation} \label{eq_4_f()}
	P^h_i \leq \frac{\alpha_{j,j} P_j^\ell}{\gamma_0^\ell a_{i,j}} \left(\frac{e^{-\frac{\gamma_0^\ell P_N}{P_j^\ell a_{j,j}}}}{1-P_o} -1\right) \triangleq f \left(P_j^\ell\right), 
\end{equation}
where $\alpha_{j,j}$ represents the links between the LCU-LCU pair given by $ \alpha_{j,j} = L_p  D^{-\phi}_{j,j} \chi_{j,j}$, where $D_{j,j}$ is the distance between two LCUs and $\chi_{j,j}$ is the respective random shadowing gain. Also, $\alpha_{i,j}$ represents the links between the $i$th HCU to $j$th LCU given by $\alpha_{i, j} =  L_p D^{-\phi}_{i,j} \chi_{i,j}$, where $D^{-\phi}_{i,j}$ is the distance between the $i$th HCU and the $j$th LCU and the corresponding shadowing gain is represented by $\chi_{i,j}$.

We define $P_{h,m}^\ell$ as the maximum LCU transmit power that enables HCU to use its maximum transmit power without violating the LCU's reliability constraint. It is given by $P_{h,m}^\ell = f^{-1} (P_{m}^h)$ and can be found using a bisection search on the monotonically increasing function $f (\cdot)$. The monotonicity of $f (\cdot)$ is evident from  \eqref{eq_4_f()} and Fig. \ref{fig_feasible} (in Appendix B). Similarly, define $P_{\ell,m}^h$ as the maximum HCU transmit power that can be used when LCU uses its maximum transmit power while satisfying the LCU's reliability constraint. It is given by $P_{\ell,m}^h = f (P_{m}^\ell)$ and can be directly obtained from \eqref{eq_4_f()}.

Consequently, the optimal power allocation solution to \eqref{eq_opt_2} can be derived for an HCU-LCU pair, aiming to maximize the capacity of an HCU while guaranteeing reliability for its spectrum-shared LCU, as given below (see Appendix~B)
\begin{equation} \label{eq_powers}
	P_i^{h^*} = \text{min} (P_{m}^h, P_{\ell,m}^h), \
	P_j^{\ell^*} = \text{min} (P_{m}^\ell, P_{h,m}^\ell).
\end{equation}
It is noted that \eqref{eq_powers} can be used to optimize power allocation for all HCU-LCU pairs. Next, we optimize the pair-matching process for spectrum sharing by considering the given objective function and constraints.

\subsection{HCU-LCU Pair Matching}
Having determined the optimal power for all HCU-LCU pairs, we now filter out any pair that fails to meet the minimum required capacity for the HCU, as specified in \eqref{eq_opt_1b}, even when \eqref{eq_powers} is applied.
The capacity achieved by the $i$th HCU when sharing the spectrum with the $j$th LCU is defined by $C_{i,j}^h (P_i^h , P_j^\ell) \triangleq C_{i,R}^h + C_{i,H}^h$ and is expressed after simplification as (see Appendix C for proof)
\begin{equation} \label{eq_cap_under_shared_spec}
	C_{i,j}^h (P_i^h , P_j^\ell) = \frac{\rho}{(\rho - \eta )\ln2}
	\left[ e^{\frac{1}{\rho}}
	E_1 \left(\frac{1}{\rho}\right) -
	e^{\frac{1}{\eta}}
	E_1 \left(\frac{1}{\eta}\right) \right],
\end{equation}
where $\rho=(P_i^h \alpha_{i,c})/P_N$,  $\eta=(P_j^\ell \alpha_{j,c})/P_N$, 	$\alpha_{i,c}=\alpha_{i,R}+\alpha_{i,H}$ and $\alpha_{j,c}=\alpha_{j,R}+\alpha_{j,H}$. $E_1(x) = \int_{x}^{\infty}(e^{-t} / t) dt$, where $E_1(x)$ is the first-order exponential integral function \cite{AM:13:IEEE_J_COML}.

By substituting \eqref{eq_powers} into \eqref{eq_cap_under_shared_spec}, we obtain the maximum capacity, denoted by $C^*_{i,j}$, for the $i$th HCU that shares its spectrum with the $j$th LCU. If $C^*_{i,j}$ falls below $C^h_0$, this particular HCU-LCU pair is deemed infeasible, implying that HCU cannot meet the minimum capacity requirement under this scenario. Therefore, we set $C^*_{i,j}=-\infty$, i.e.,
\begin{equation}
	C^*_{i,j}= 
	\begin{cases}
		C_{i,j}^h (P_i^{h^*} , P_j^{\ell^*}),& \text{if } C_{i,j}^h (P_i^{h^*} , P_j^{\ell^*}) \geq C^h_0,\\
		-\infty,              & \text{otherwise.}
	\end{cases}
\end{equation}
This equation can be translated into the following optimization problem, which the HM can solve efficiently in polynomial time \cite{IM:07:B_E}:
\begin{subequations}
	\begin{align}
		\max_{\substack{\{\mu_{i,j}\}
		}} \quad & \sum_{i\in I} \sum_{j \in J} \mu_{i,j} C^*_{i,j} \\
		\text{s.t.  } & \eqref{eq_opt_1f} - \eqref{eq_opt_1h}.
	\end{align}
\end{subequations}
Based on the formulation above, we introduce Algorithm \ref{alg1} to solve \eqref{eq_opt_1}. 

\begin{algorithm}[t]
	\small
	\caption{Solution to \eqref{eq_opt_1}}
	\begin{algorithmic}[1] \label{alg1}
		\STATE \textbf{Input:}  $C^h_0, P_{m}^h, P_{\ell,m}^h,  P_{m}^\ell, P_{h,m}^\ell$
		\STATE \textbf{Output:} Optimal spectrum sharing and power allocation patterns $\{\mu_{i,j}^*\}, \{P_i^{h^*}\}, \{P_j^{\ell^*}\}$.
		\FOR{$i = 1 : I$}
		\FOR{$j = 1 : J$}
		\STATE Using \eqref{eq_powers}, determine the optimal power allocation $(P_i^{h^*}, P_j^{\ell^*})$ for the specific HCU-LCU pair under consideration.
		\STATE Substitute $(P_i^{h^*}, P_j^{\ell^*})$ into \eqref{eq_cap_under_shared_spec} to find $C^*_{i,j}$.
		\IF{$C^*_{i,j} < C^h_0$}
		\STATE $C_{i,j}^* = -\infty$
		\ENDIF
		\ENDFOR
		\ENDFOR
		\STATE Following the HM, determine the optimal spectrum sharing pattern \{$\mu_{i,j}^*$\} based on $\{C_{i,j}^*\}$.
		\STATE Obtain \{$\mu_{i,j}^*$\} and the associated  $\{(P_i^{h^*}, P_j^{\ell^*})\}$.
	\end{algorithmic}
\end{algorithm}

\textit{Complexity}: Given an accuracy requirement of $\epsilon$, the bisection search requires $\log(1/\epsilon)$ iterations for \eqref{eq_powers}, implying that a complexity of $\mathcal{O}(JI\log(1/\epsilon))$ is incurred in total to solve \eqref{eq_powers} across all HCU-LCU spectrum-sharing pairs. The HM accomplishes pair matching in the complexity of $\mathcal{O}(I^3)$ time if $I \geq J$. Hence, the overall computational complexity becomes $\mathcal{O}(JI\log(1/\epsilon) + I^3)$.


\section{Maximizing the Minimum HCU Capacity} \label{sec_min_cap_max}
The capacity optimization in the previous section enhances overall network throughput but overlooks fairness, potentially leading to poor performance for HCUs with bad channel conditions. This section focuses on maximizing the minimum capacity achieved by all HCUs, to provide constant QoS for all HCUs. Thus, our optimization problem is presented as follows:
\begin{equation}\label{eq_min_cap_max}
	\begin{aligned} 
		\max_{\substack{\{\mu_i,j\},  \{P^{h}_{i}\}, \{P^{\ell}_{j}\}
		}} \quad & \min_{i\in I} C_i^h  \\
		\text{s.t. } 	& \eqref{eq_opt_1b} - \eqref{eq_opt_1h}
	\end{aligned}
\end{equation}
The above equation, formulated as a max-min optimization, guarantees to achieve the Pareto optimal, as noted in \cite{EAMB:14:IEEE_M_SP} and \cite{GYLY:15:IEEE_J_JSAC}, ensuring that no capacity of HCU can be further improved without compromising the capacity of other HCUs.


\begin{algorithm}[t]
	\small
	\caption{Solution to \eqref{eq_min_cap_max}}
	\begin{algorithmic}[1]\label{alg2}
		\STATE \textbf{Input:}   $\{C^*_{i,j}\}$ and $\{(P_i^{h^*}, P_j^{\ell^*})\}$ from Algorithm \ref{alg1}
		\STATE \textbf{Output:} Optimal spectrum sharing and power allocation patterns $\{\mu_{i,j}^*\}, \{P_i^{h^*}\}, \{P_j^{\ell^*}\}$.
		\STATE	Set $k=1$
		\STATE	Set $m=JI$
		\STATE	Rearrange elements of $\{C^*_{i,j}\}$ from highest to lowest order and keep in a vector $\textbf{g}$.
		\WHILE{$(m-1)> 1$}
		\STATE $l= (k+m)/2$
		\STATE $ \textbf{L} = \textbf{0}_{I \times J}$
		\FOR{$i = 1 : I$}
		\FOR{$j = 1 : J$}
		\IF{$C^*_{i,j} < \textbf{g}_l$}
		\STATE {$\textbf{L}_{i,j} = 1$}
		\ELSE
		\STATE {$\textbf{L}_{i,j} = 0$}
		\ENDIF
		\ENDFOR
		\ENDFOR
		\STATE Based on $\textbf{L}$, use the HM to determine the assignment set $\textbf{D}$ and the least aggregated cost value $\delta$.
		\IF {$\delta>0$}
		\STATE {$m=l$}
		\ELSE
		\STATE $k=l$
		\STATE $\{\mu_{i,j}^*\} = \textbf{D}$
		\ENDIF
		\ENDWHILE
		\STATE Get $\{\mu_{i,j}^*\}$ and the associated $\{(P_i^{h^*}, P_j^{\ell^*})\}$.
		
	\end{algorithmic}
	
\end{algorithm}

\subsection{Designing Resource Allocation Strategy} 
Our resource allocation problem formulated in  \eqref{eq_min_cap_max} is solved by utilizing the optimal power values listed in  \eqref{eq_powers}  for each HCU-LCU pair and the HCU's capacity derived in \eqref{eq_cap_under_shared_spec}. Solving the problem this way, we assume that interference exists between each HCU-LCU pair only. 
Consequently, \eqref{eq_min_cap_max} is transformed into the following simple form:
\begin{subequations} \label{eq_max_min}
	\begin{align}
		\max_{\{\mu_{i,j}\}} \quad & \min_{i\in I} \sum_{j \in J} \mu_{i,j} C_{i,j}^* \\
		\text{s.t.  } & \eqref{eq_opt_1f} - \eqref{eq_opt_1h}  \label{eq_16b}
	\end{align}
\end{subequations}

To solve the above optimization problem, we propose Algorithm \ref{alg2}. The algorithm leverages the HM, which has polynomial time complexity, thereby ensuring efficient computation. The algorithm consists of two key steps.
First, a minimum capacity check is conducted to determine, within polynomial time, whether an arbitrarily defined threshold $C_{min}$ exceeds the desired minimum optimal capacity, where $C_{min}$ is a conceptual variable representing the minimum capacity.  Its discrete value is denoted by $\textbf{g}_l$ in the algorithm.
The algorithm proceeds as follows.
\begin{itemize} [leftmargin=10pt, labelsep=5pt]
	\item A zero matrix, denoted by \textbf{L}, with dimensions $I \times J$ is initialized, which is used to store the results of the minimum capacity check. 
	\item Each element of the capacity matrix $\{C^*_{i,j}\}$, obtained from Algorithm \ref{alg1}, is examined. If the value of any element is lower than $C_{min}$, the corresponding entry in \textbf{L} is set to one. Conversely, if any element has a value greater than or equal to $C_{min}$, the entry is set to zero. In other words, for any $i$ and $j$, it can be expressed as
	\begin{equation}
		\textbf{L}_{i,j} = 
		\begin{cases} 
			1, & \text{if } C^*_{i,j} < C_{min}, \\
			0, & \text{otherwise.}
		\end{cases}.
	\end{equation}
	
	\item The HM is applied to $\textbf{L}$ and the lowest total cost, $\delta$, is found. This cost is the sum of all non-zero assigned elements. If $\delta=0$, this implies that all elements under this assignment are greater than $C_{min}$, or $C_{min}$ does not exceed the desired optimal capacity. Conversely, if $\delta > 0$, then no feasible assignment exists meaning that all the assigned elements are not greater than $C_{min}$, or stated otherwise, $C_{min}$ exceeds the desired optimal minimum capacity.
\end{itemize}

In the second step, all $J I$ elements of $\{C^*_{i,j}\}$ are sorted in order. Then, the position of the optimal minimum capacity is searched for using the bisection search method described above. Once the bisection search has converged, the HM is applied to yield the spectrum sharing assignment.

\textit{Complexity:} 
The generation of $\{C^*_{i,j}\}$ has $\mathcal{O}(JI\log(1/\epsilon))$ complexity, whereas sorting all the $JI$ capacity values in ascending order to facilitate the bisection search has $\mathcal{O}(JI\log(JI))$ complexity. 
Additionally, the HM has complexity of $\mathcal{O}(I^3)$ (assuming $I \geq J$), and the bisection search over $JI$ elements requires $\mathcal{O}(\log I)$ iterations, leading to $\mathcal{O}(I^3 \log I)$ computations. The overall complexity amounts to
$\mathcal{O}\left(JI\log\left(1/\epsilon\right) + JI\log(JI) + I^3\log I\right)$.

\renewcommand{\arraystretch}{1.15}
\begin{table}[t]
	\centering
	\caption{Simulation parameters based on \cite{SSS:20:IEEE_J_IOT, WOI:17:IEEE_C_VTC, TYYWA:20:IEEE_J_WC, YFAR:21:EEE_C_VTC, FMC:23:IEEE_J_TNSM,SYLR:19:IEEE_J_VT}.}
	\label{tab_sim_params}
	\begin{tabular}{|p{5.3cm}|p{2.25cm}|}
		\hline
		
		\hline
		\textbf{Parameter} & \textbf{Value} \\
		\hline
		Carrier frequency & 2 GHz \\ \hline
		Channel bandwidth & 10 MHz \\ \hline
		RBS antenna height with gain & 20 m with 8 dBi 			  \\ \hline
		RBS receiver noise figure  		&  5 dB \\ \hline
		Noise power	($P_N$)		&  -114 dBm \\ \hline
		UAVs height 	&  0.1 km \\ \hline
		UAVs antenna height	and antenna gain	& 0.2 m and 3 dBi \\ \hline
		UAVs receiver noise figure			&  9 dB\\ \hline
		Distance between RBS and UAVs 			&   25 m\\ \hline
		The SINR ($\gamma_0^\ell$) and reliability ($P_o$) constraints &  5 dB and $10^{-3}$ \\ \hline
		
		The number of HCU and LCU	 &  20 each \\ \hline
		Max. transmit powers for HCU ($P_{m}^h$)	 and LCU  ($P_{m}^\ell$)	  &  16 dbm and 22 dbm \\ \hline
		$\chi_{i,R}$ 	   & 8 dB  \\ \hline
		$\chi_{i,j}$ and $\chi_{i,H}$ 	    & 3 dB each  \\ \hline
		UAV's trajectory 	      &  Straight line \\ \hline
		HAP altitude 	       & 17 km  \\ \hline
		HAP noise figure 	        &  3 dB  \\ \hline
		HAP antenna again        & 8 dBi  \\ \hline
		UAVs speed & 70 km/h \\ \hline
		Accuracy threshold in the Bisection search ($\epsilon$)  	& $10^{-5}$
		\\ \hline
		Min. capacity requirements of HCUs ($C_0^h$)	& 0.5 bps/Hz \\ \hline
		
		\hline
	\end{tabular}
\end{table}

\textcolor{red}{\textit{Convergence}: Algorithm \ref{alg1} has finite-termination and doesn't include an indefinite iterative loop. 
	Its only numerical iteration arises in computing $P^{\ell}_{h,m}=f^{-1}(P^h_m)$ through bisection on $f(\cdot)$. 
	Since $f(\cdot)$ is monotonically increasing over the feasible interval, bisection converges to an $\epsilon$-accurate solution. 
	Moreover, after evaluating all $JI$ pairs, Algorithm \ref{alg1} runs the HM, which deterministically returns the optimal one-to-one assignment in polynomial time.
	Hence, Algorithm \ref{alg1} terminates after a finite number of steps and yields a well-defined solution.
	Algorithm \ref{alg2} performs a bisection search over the discrete set of candidate capacities obtained from $\{C_{i,j}^*\}$.
	At each step, it checks the feasibility of achieving a candidate minimum-capacity threshold by constructing the binary matrix $L$ and solving an assignment problem through the HM. 
	Because the search space contains $JI$ discrete candidates, the bisection steps terminate in at most $\mathcal{O}(\log(JI))$ steps, and each step completes in polynomial time. Therefore, Algorithm \ref{alg2} also terminates in finite steps.}

\section{Results and Discussion} \label{sec_results}
In this section, we evaluate the proposed resource allocation algorithms through simulations using the parameters and configurations listed in Table \ref{tab_sim_params}. Further details are provided below.
\textcolor{blue}{We consider an integrated TN--NTN scenario comprising one RBS, one HAP node, and a set of $I$ HCUs and $J$ LCUs. Unless stated otherwise, the number of HCUs and LCUs is fixed at 20 each, corresponding to a moderately dense aerial network scenario.
	The simulation area is a square region of size $2~\mathrm{km} \times 2~\mathrm{km}$.  
	The RBS is placed at the center, while the HAP is positioned vertically above the center at an altitude of 17 km.  
	HCUs are randomly deployed at an altitude of 0.1 km.  
	The positions of UAVs follow a spatial Poisson process \cite{XCFH:21:IEEE_J_VT}. 
	The density of UAVs is non-uniform and varies with their speed. On average, the distance between UAVs is given by two seconds multiplied by their absolute speed. All UAVs follow straight-line mobility.
	Furthermore, the $I$ HCUs and $J$ pairs of LCUs are randomly selected within the simulated area, and LCU pairs are formed only between neighboring UAVs.
	The log-distance path loss model with log-normal shadowing and Rayleigh fading is assumed, as in \cite{YJHZCL:23:J_COMNET}. 
	The total available bandwidth is 10 MHz and the carrier frequency is 2 GHz.  
	The transmit powers are 22 dBm for HCUs and 16 dBm for LCUs.}

All the results are averaged over at least $10^3$ channel realizations. For brevity, Algorithm \ref{alg1} and Algorithm \ref{alg2} are labeled as ``Algo 1" and ``Algo 2", respectively, in the figures showing the results.

%
\begin{figure}[t]
	\centering
	\subfigure[The sum capacity of HCUs]{\includegraphics[width=0.9\columnwidth]{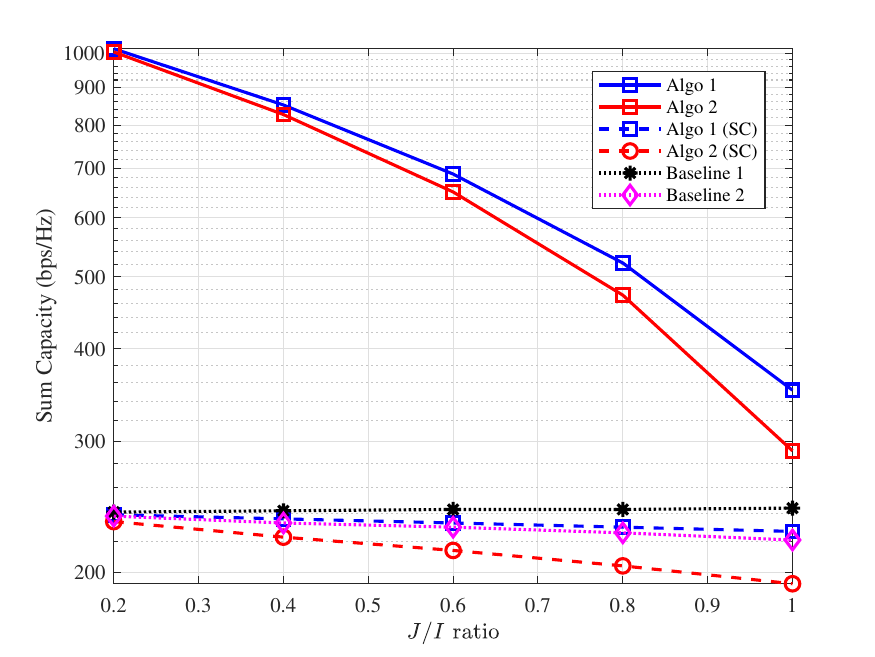}}
	\hfil
	\subfigure[The minimum capacity of HCUs]{\includegraphics[width=0.9\columnwidth]{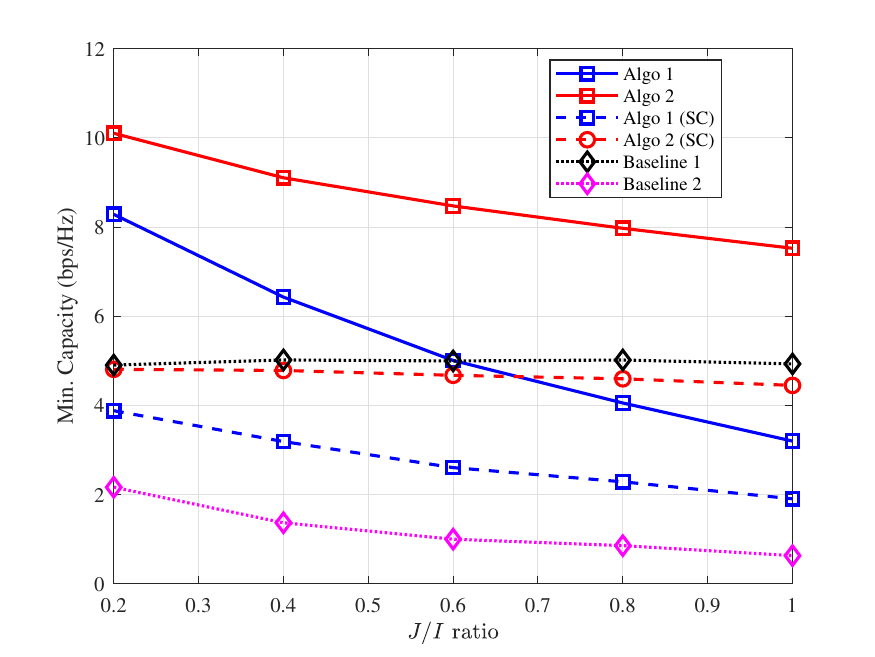}}
	\hfil
	\caption{The capacity performance comparison with different $J/I$ ratios.}
	\label{fig_cap_comparison}
\end{figure}

\subsection{Capacity Comparison}
Figs. \ref{fig_cap_comparison}(a) and (b) show the performance of our proposed algorithms, as a function of the $J/I$ ratio, against no resource sharing in \cite{JJMY:22:IEEE_J_VT} (i.e., Baseline 1) and greedy resource sharing in \cite{KHYA:21:IEEE_J_CCN} and \cite{MMNRHE:24:IEEE_J_COM} (i.e., Baseline 2).
Baseline 1 allocates orthogonal spectrum bands to each UAV and does not allow any other UAV to reuse these bands. Thus, Baseline 1 serves as the upper bound for performance gain under the considered spectrum-sharing framework.
In Baseline 2, spectrum sharing is performed, and HCU-LCU pairings are decided in the form of an incumbent and secondary user pair using a greedy and heuristic approach aimed at maximizing the sum capacity of the HCUs. Additionally, none of the baseline algorithms utilize MC. 
To highlight the performance gain due to MC, the single connectivity (SC) case is also presented in the results.






In Fig. \ref{fig_cap_comparison}(a), as the $J/I$ ratio increases, the sum capacity generally declines for all competing schemes except Baseline 1, which remains constant. The proposed algorithms under MC substantially outperform their SC counterparts in terms of capacity, demonstrating the advantage of exploiting multiple links simultaneously. Moreover, Algorithm \ref{alg1} achieves the highest sum capacity across all $J/I$ values due to its design goal of maximizing total throughput while ensuring LCU reliability. Algorithm \ref{alg1} consistently outperforms Algorithm \ref{alg2} and Baseline 2, maintaining higher capacity across all $J/I$ ratios. Algorithm \ref{alg2} shows the second-highest yet steepest decline in performance, indicating its higher sensitivity to increased interference. Baseline 1 shows flat performances across all $J/I$ ratios and outperforms Baseline 2 and both the proposed algorithms under SC. This is because Baseline 1 does not employ spectrum sharing. Baseline 2 performs better compared to Algorithm 2; however, its performance remains inferior to that of Algorithm 1 across all $J/I$ ratios.

In Fig. \ref{fig_cap_comparison}(b), we observe that Algorithm \ref{alg2} achieves the highest minimum capacity across all $J/I$ values, outperforming all other schemes. This is because  Algorithm \ref{alg2} is designed to maximize the minimum capacity across all HCUs, thereby promoting fairness in resource allocation. Notably, Algorithm \ref{alg2} maintains relatively stable performance even when interference increases due to a higher number of LCUs.
Algorithm \ref{alg1} exhibits a declining trend in the minimum capacity as the $J/I$ ratio increases. While it outperforms all other schemes, its performance falls below that of Baseline 1, when $J/I > 0.6$, and it falls behind Algorithm \ref{alg2} (SC) for $J/I>0.7$.
This reflects its focus on maximizing the sum capacity rather than ensuring fairness.
Baseline 1 achieves the most stable minimum capacity, remaining around 5 bps/Hz regardless of $J/I$ ratio. Algorithm \ref{alg2} (SC) performs slightly below Baseline 1, but maintains relatively stable performance, suggesting robustness to interference in terms of fairness. In contrast, Algorithm \ref{alg1} (SC) and Baseline 2 show significant degradation in the minimum capacity as the $J/I$ ratio increases. Baseline 2 performs the worst overall. This indicates that while Algorithm \ref{alg1} and Baseline 1 enhance the sum capacity, they compromise user fairness or worst-case performance under interference.



\begin{figure}
	\centering
	\subfigure[The instantaneous sum capacity of HCUs]
	{\includegraphics[width=0.9\linewidth]{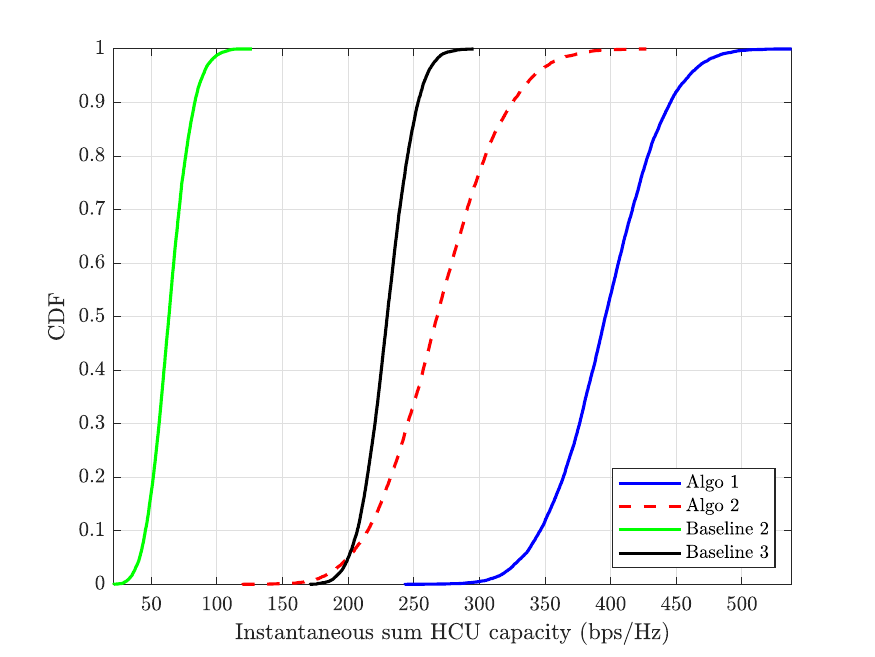}}
	\hfill
	\subfigure[The instantaneous SINR of LCUs]
	{\includegraphics[width=0.9\linewidth]{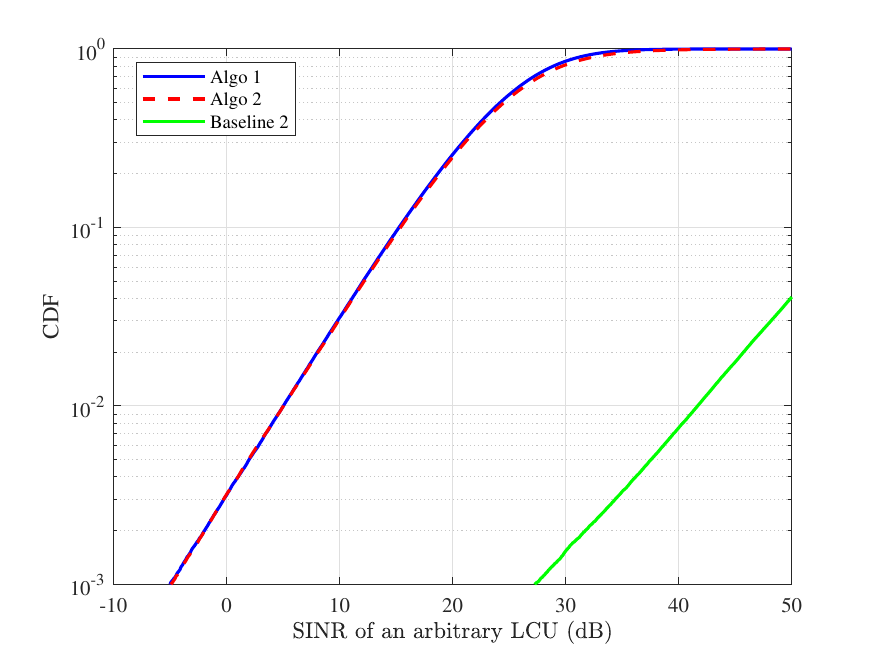}} 		
	\caption{The CDFs of the instantaneous system performances, when $P_{m}^{l} = P_{m}^{h} = 22$ dBm and $P_o = 0.01$.}
	\label{fig_cdf}
\end{figure}


\subsection{Instantaneous System Performance Comparison}
In Fig. \ref{fig_cdf}(a), we compare the cumulative distribution functions (CDFs) of the instantaneous sum capacity\footnote{Instantaneous sum capacity is computed for a single realization of all random channel variables during one simulation instance.}. of HCU obtained by our proposed algorithms, Baseline 2, and no resource sharing (labeled as Baseline 3) \cite{JJMY:22:IEEE_J_VT}. Baseline 2 uses HCU-LCU pairing while forcing both the HCU and its paired LCU to transmit at their maximum power. This comparison illustrates the gains achieved by our proposed optimal power control, followed by the proposed algorithms compared to a simpler maximum power approach with good pairing, followed by Algorithm 3. Similarly, in Baseline 3, only HCUs are active and transmit at maximum power, and there is no UAV-UAV communication reusing the spectrum. Therefore, LCUs effectively cause no interference to HCUs, and HCUs cause no interference to LCUs operating on this spectrum (as LCUs are considered inactive on these HCU resources). This baseline demonstrates the sum capacity gain from UAV-UAV spectrum sharing and serves as the performance upper bound.

It can be seen that both Algorithms \ref{alg1} and \ref{alg2} outperform both baselines. Baseline 2 saturates at a low capacity range, which indicates its limitation in supporting high-throughput applications. Although Baseline 3 improves upon Baseline 2 by utilizing cognitive radio-based techniques, it performs worse than the proposed algorithms due to its limited spectrum reuse and lack of fairness considerations.
The steepness in Baselines 2 and 3 reveals their inefficient resource allocation strategies with low adaptability to channel variations. In contrast, the smoother slopes of our proposed algorithms indicate more flexible and robust designs.


In Fig. \ref{fig_cdf}(b), the CDFs of our proposed algorithms and Baseline 2 are plotted as functions of the instantaneous SINR of an arbitrarily chosen LCU. It is worth noting that in Baseline 3 \cite{JJMY:22:IEEE_J_VT}, LCUs do not operate on the shared spectrum; hence, their SINRs cannot be plotted in Fig. \ref{fig_cdf}(b). We also observe that LCUs in Baseline 2 cannot achieve the targeted SINR threshold of 5 dB with the given $P_o$, implying that a substantial portion of LCUs experience higher outage probabilities. On the other hand, both the proposed algorithms achieve the required threshold of 5 dB at the given $P_o$, ensuring that a large number of LCUs meet the minimum SINR requirements with a very low outage probability. The overlap between the proposed algorithms indicates that they both provide equally robust reliability performance for LCUs.


\begin{figure}
	\centering
	\subfigure[The sum capacity of HCUs]
	{\includegraphics[width=0.9\linewidth]{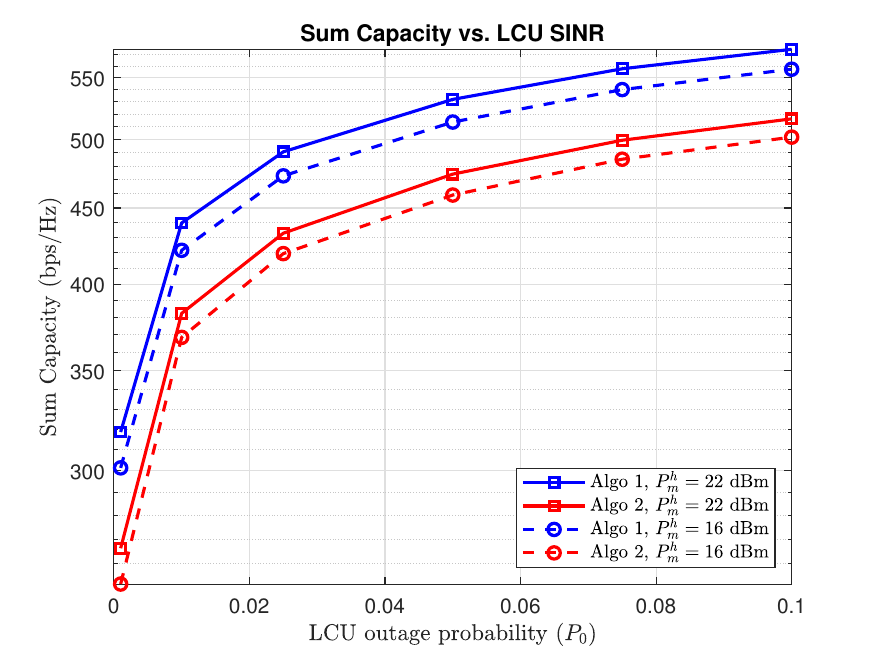}}
	\hfill
	\subfigure[The minimum capacity of HCUs]
	{\includegraphics[width=0.9\linewidth]{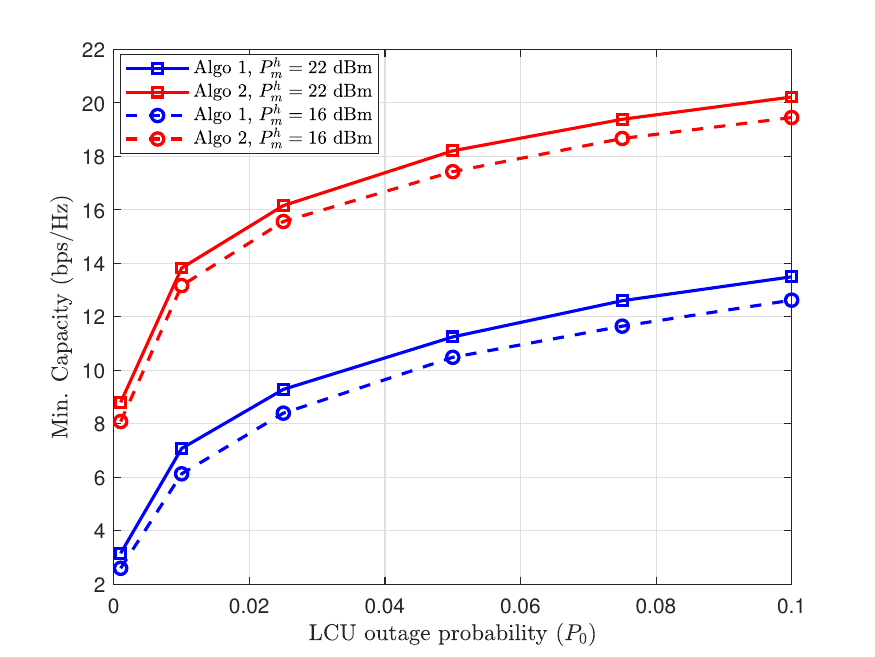}} 		
	\caption{The impact of the LCU outage probability.}
	\label{fig_cap_vs_p0}
\end{figure}

\subsection{Impact of LCU Outage Probability on Capacity of HCUs}

Figs. \ref{fig_cap_vs_p0}(a) and (b) show the sum capacity and the minimum capacity of HCUs plotted as functions of LCU outage probability $P_o$, respectively. The figures indicate that, for a given transmission power, increasing $P_o$ increases the capacities. Moreover, Algorithm \ref{alg1} outperforms Algorithm \ref{alg2} in Fig. \ref{fig_cap_vs_p0}(a) while the reverse is true in Fig. \ref{fig_cap_vs_p0}(b). Additionally, increasing the maximum transmit power results in higher sum capacity and minimum capacity for both algorithms. 
The gradual increase in the capacities of HCUs under both algorithms as $P_o$ increases is due to the fact that a higher $P_o$ means greater tolerance of LCUs towards HCU's interference, which prompts HCUs to transmit at increased power levels. 
Algorithm \ref{alg1} yields improved sum capacity compared to its counterpart, as shown in Fig. \ref{fig_cap_vs_p0}(a), due to its design objective of maximizing the sum capacity. On the other hand, Algorithm \ref{alg2} ensures better performance for the worst-case scenario, as shown in Fig. \ref{fig_cap_vs_p0}(b), because it maximizes the minimum capacity. More precisely, Algorithm \ref{alg1} prioritizes sum capacities at the expense of fairness, thereby sacrificing the capacity of users with poor channel conditions, which leads to lower minimum capacity. In contrast, Algorithm \ref{alg2} focuses on fairness, even if it means compromising the total capacity of the system.


\begin{figure}
	\centering
	\subfigure[The sum capacity of HCUs]
	{\includegraphics[width=0.9\linewidth]{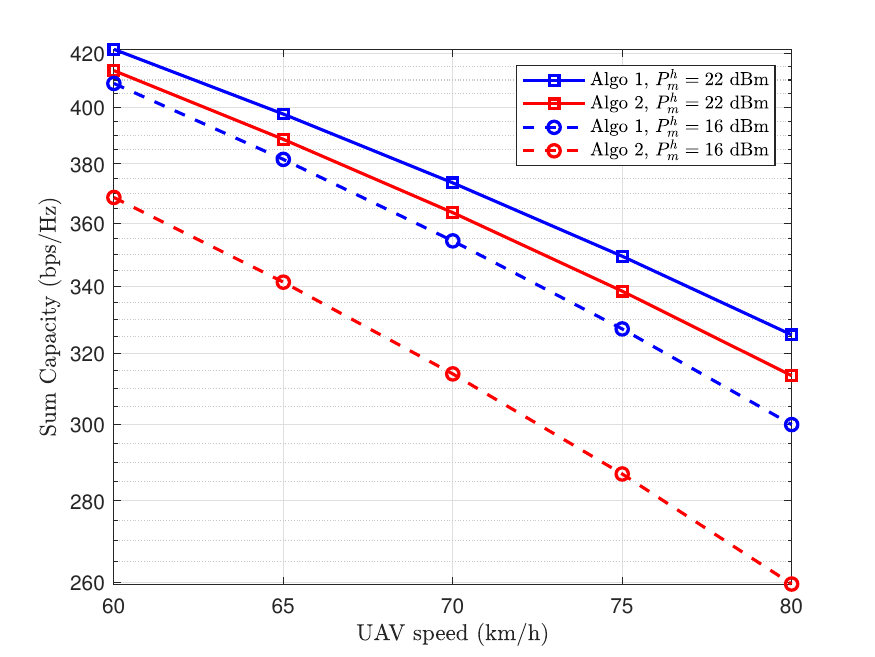}}
	\hfill
	\subfigure[The minimum capacity of HCUs]
	{\includegraphics[width=0.9\linewidth]{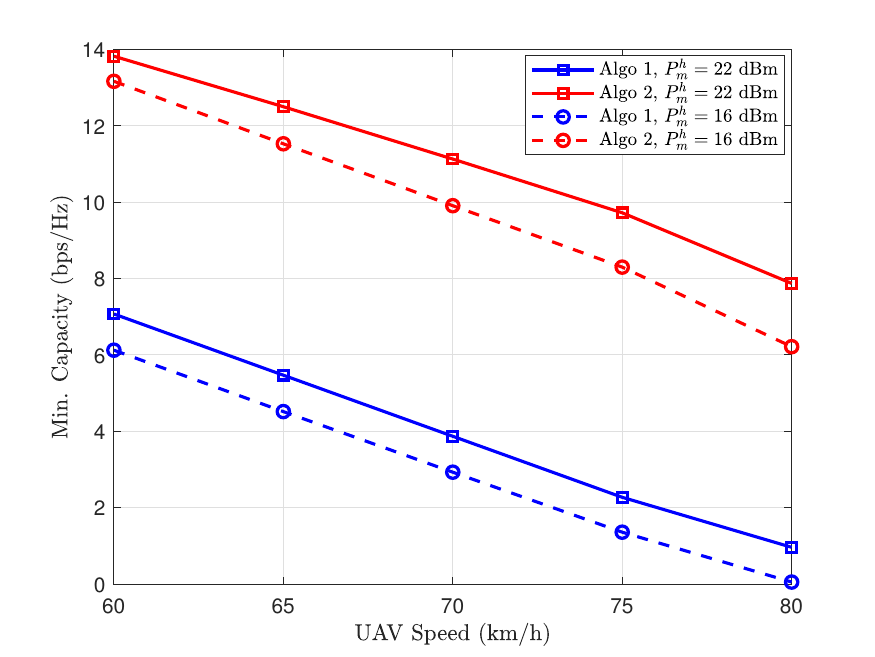}}
	\caption{The impact of the UAV speed.}
	\label{fig_cap_vs_speed}
\end{figure}

\subsection{Impact of the UAV Speed}
Figs. \ref{fig_cap_vs_speed}(a) and (b) illustrate the sum and minimum capacity performance of HCUs as functions of the UAV speed, respectively. It can be witnessed that an increase in UAV speed results in a larger average inter-UAV distance and sparser traffic, as defined in the simulation setup. This increased distance weakens the direct UAV-UAV channel gain $\alpha_{j,j}$, which makes it difficult to satisfy the LCU reliability constraint, necessitating reduced HCU's transmission power to protect the LCU link. This reduction in transmission power lowers the HCU capacity. Hence, the figures display decreasing capacity trends as vehicle speed increases. Moreover, as shown in Fig. \ref{fig_cap_vs_speed}(a), the sum capacity in both algorithms shows a comparatively stable response to changes in maximum transmit power. However, in the case of minimum capacity, Algorithm \ref{alg1} displays a comparatively lower impact on changes in the maximum transmission power, as shown in Fig. \ref{fig_cap_vs_speed}(b).


\begin{figure}
	\centering
	\subfigure[The sum capacity of HCUs]
	{\includegraphics[width=0.9\linewidth]{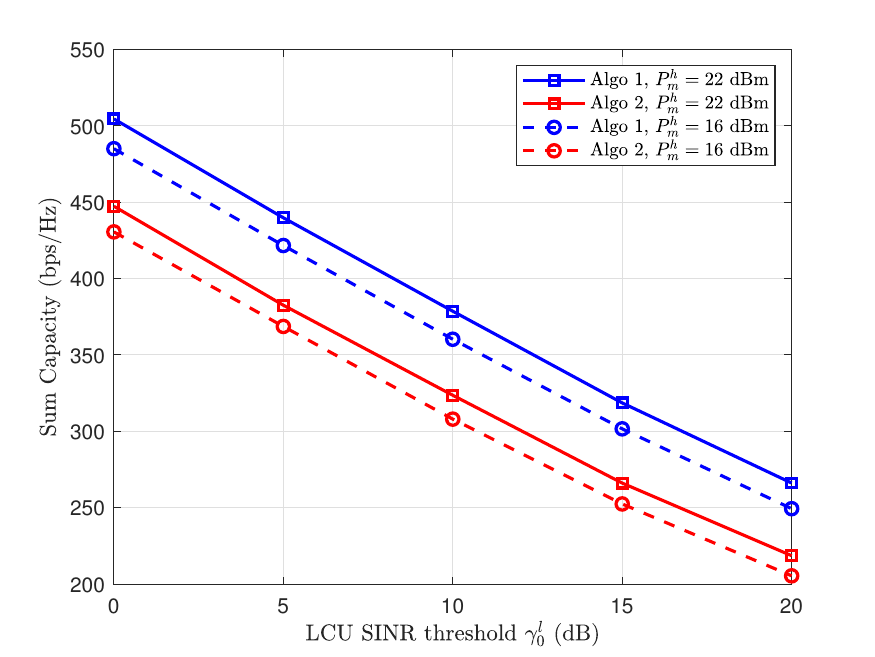}}
	\hfill
	\subfigure[The minimum capacity of HCUs]
	{\includegraphics[width=0.9\linewidth]{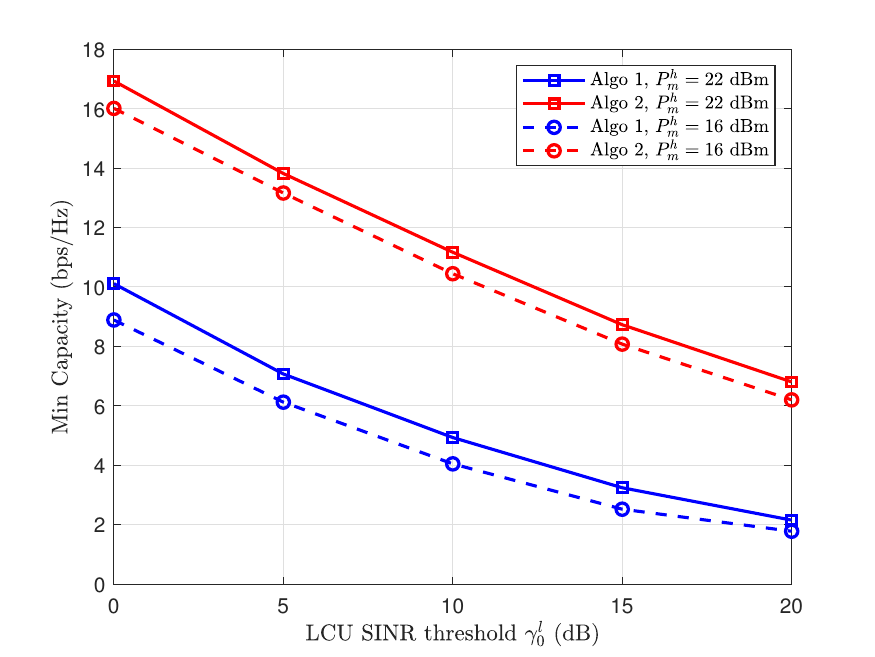}}
	\caption{The impact of the SINR of LCUs.}
	\label{fig_cap_vs_sinr}
\end{figure}

\subsection{Impact of the SINR of LCUs}

In Figs. \ref{fig_cap_vs_sinr}(a) and (b), we investigate how the sum and minimum ESs of HCUs vary with the LCU SINR threshold $\gamma_0^\ell$. It is observed that a higher value of $\gamma_0^\ell$ makes the LCU reliability constraint stricter, as characterized in \eqref{eq_4_f()}. Consequently, LCUs become less tolerant to interference from HCUs, which reduces the feasible transmission power range for HCUs. This prompts HCUs to transmit with lower power, thereby decreasing their capacity, both in the sum and minimum capacity sense. Therefore, Figs. \ref{fig_cap_vs_sinr}(a) and (b) show the decreasing trends of capacity as $\gamma_0^\ell$ increases. Moreover, as shown in Fig. \ref{fig_cap_vs_sinr}(a), a change in the transmission power exhibits nearly constant impact on the sum capacity as $\gamma_0^\ell$ increases. Contrarily, changing the maximum transmission power disproportionately impacts the minimum capacity, which diminishes swiftly with higher values of $\gamma_0^\ell$, as shown in Fig. \ref{fig_cap_vs_sinr}(b).


\begin{figure}
	\centering
	\subfigure[The sum capacity of HCUs]
	{\includegraphics[width=0.9\linewidth]{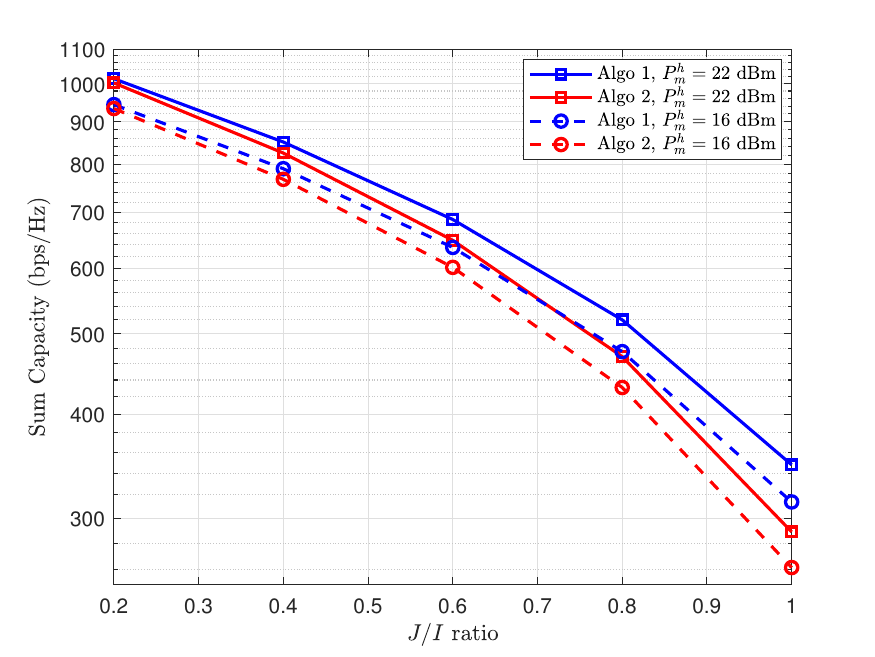}}
	\hfill
	\subfigure[The minimum capacity of HCUs]
	{\includegraphics[width=0.9\linewidth]{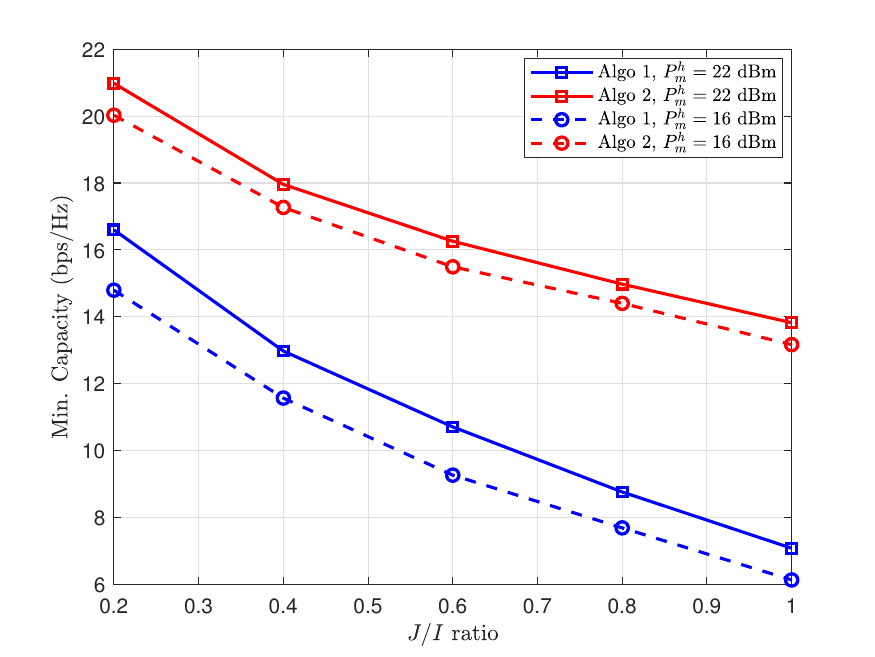}}
	\caption{The impact of the $J/I$ ratio and maximum transmit power.}
	\label{fig_cap_vs_JI_ratio}
\end{figure}

\subsection{Impact of the $J/I$ Ratio and Maximum Transmit Power}

Figs. \ref{fig_cap_vs_JI_ratio}(a) and (b) show how the active UAV-UAV links affect the quality of UAV-RBS links. It is observed that as the $J/I$ ratio or the number of UAV-UAV links increases, the overall interference in the system also increases. It is because a higher number of LCUs introduces more potential sources of interference at the RBS for HCUs. 
In Fig. \ref{fig_cap_vs_JI_ratio}(a), Algorithm \ref{alg2} responds more significantly to variations in the $J/I$ ratio and exhibits sharp degradation in sum capacity as the ratio increases. However, Algorithm \ref{alg1} degrades gracefully compared to Algorithm \ref{alg2}.
In Fig. \ref{fig_cap_vs_JI_ratio}(b), the reverse is true, and Algorithm \ref{alg2} shows more graceful degradation compared to Algorithm \ref{alg1}, maintaining higher minimum capacity. Interestingly, the minimum capacity of Algorithm \ref{alg2} is significantly more affected by changes in maximum transmit power compared to Algorithm \ref{alg1}, while the impact on the sum capacity is approximately the same in both algorithms.


\begin{figure*}
	\centering
	\setlength{\tabcolsep}{0pt} 
	\begin{tabular}{ccc}
		\includegraphics[width=0.335\linewidth]{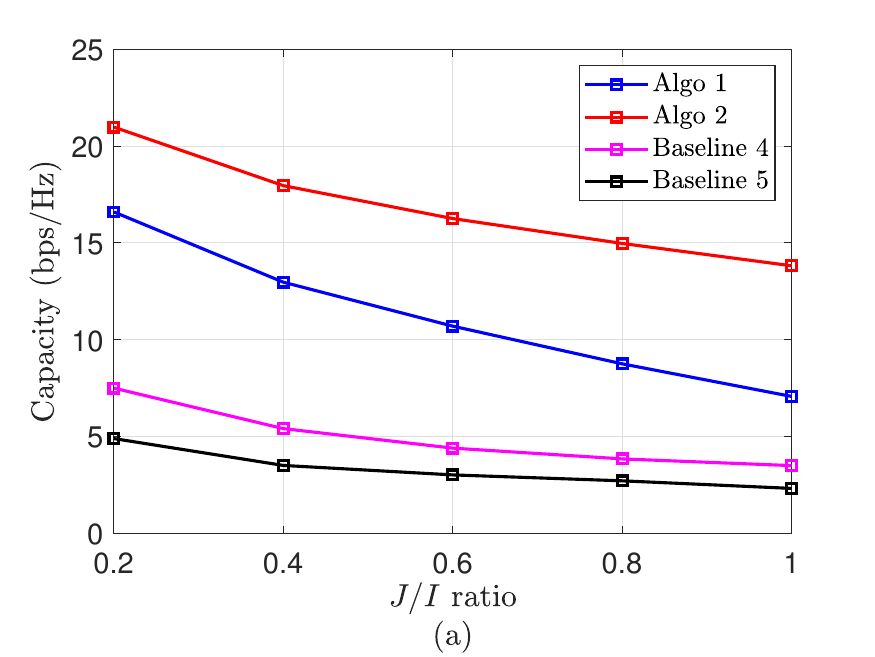} &
		\includegraphics[width=0.335\linewidth]{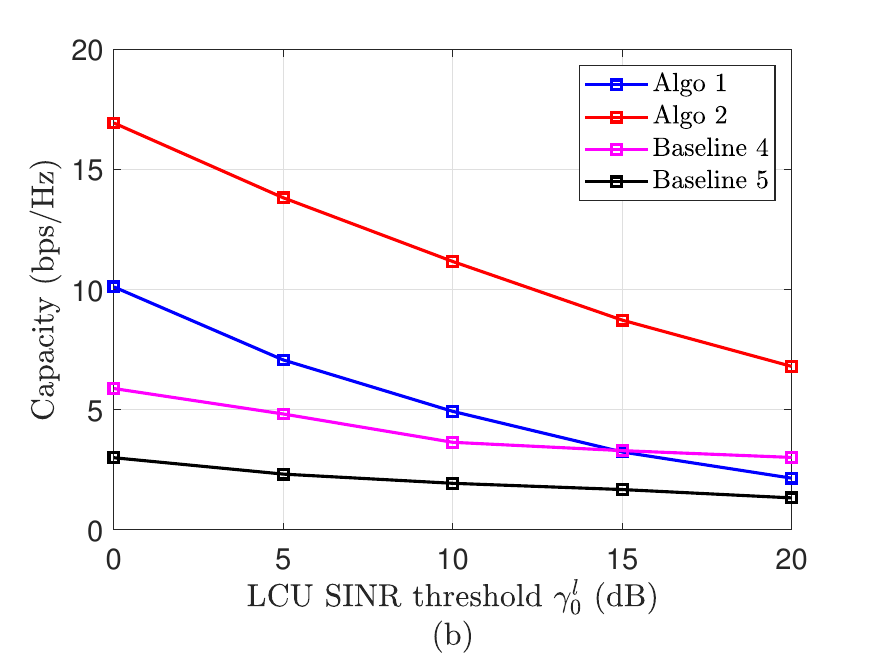} &
		\includegraphics[width=0.335\linewidth]{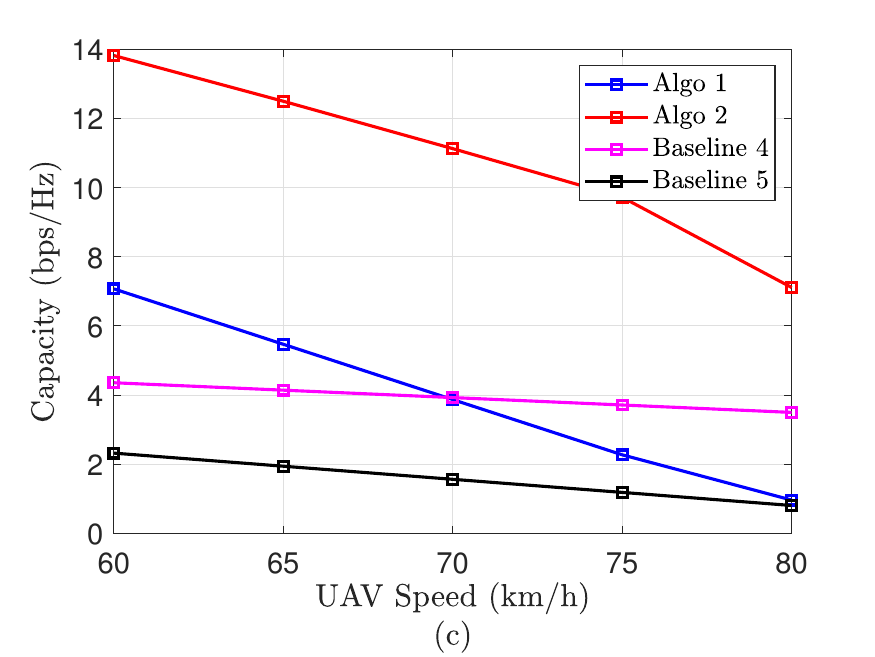}
	\end{tabular}
	\caption{Performance comparison with existing optimization schemes}
	\label{fig_new_baselines}
\end{figure*}

\textcolor{blue}{\subsection{Additional Performance Comparisons}  We further evaluate the proposed algorithms against two additional baselines called Baseline 4 and Baseline 5. 
	Baseline 4 optimizes capacity using fixed maximum transmission power and random HCU-LCU pairing. This baseline is inspired by \cite{10278101} and \cite{10999814}.
	Baseline 5 optimizes capacity using greedy spectrum-sharing, in which UAV pairs are formed sequentially based on instantaneous capacity gains. The pairing process prioritizes locally optimal links, while transmit powers are fixed at their maximum values. This baseline is derived from \cite{9170614} and \cite{10736937}. Both baselines are equipped with multiconnecity in the following evaluation for fair comparison. }

\textcolor{blue}{Fig.~8(a) shows minimum capacity as a function of  $J/I$ ratio. As seen in the figure, both the proposed algorithms outperform the baseline schemes across the given $J/I$ ratio range. Both the baseline schemes exhibit significantly lower capacities. The reason is that both the baseline schemes do not jointly optimize spectrum sharing and power allocation under reliability constraints. Moreover, Baseline 4 performs better than Baseline 5 because the latter pairs HCUs and LCUs based on instantaneous channel gains, ignoring interference coupling and fairness.}

\textcolor{blue}{In Fig.~8(b), Algo \ref{alg2} consistently achieves the highest capacity, while Algorithm \ref{alg1} also maintains a clear performance margin over the baselines, although its capacity degrades more rapidly at higher SINR thresholds. Baseline 4 and Baseline 5 exhibit substantially lower capacities at lower SINR thresholds; however, they show more graceful degradation than the proposed algorithms, with Baseline 4 surpassing Algorithm \ref{alg1} at an SINR threshold of 15. The reason is that Baseline 4 uses random pairing with fixed power allocation.}

\textcolor{blue}{In Fig. 8(c) as well, Algorithm \ref{alg2} outperforms the baseline schemes across the entire range of UAV speeds. Algorithm~\ref{alg1} also outperforms both the baselines at lower UAV speeds; however, its performance degrades quickly as the UAV speed increases. Both the baselines achieve significantly lower capacities overall; however, they exhibit more gradual performance degradation with increasing speed. Notably, Baseline 5 approaches, while Baseline 4 surpasses, Algorithm \ref{alg1} at higher UAV speeds. Overall, Fig. 8 confirms that the proposed algorithms consistently outperform existing optimization-based baselines under a wide range of network configurations.}


\section{Discussions on Practical Applications and Future Extensions}
The proposed resource sharing and allocation framework demonstrates clear advantages in capacity maximization in the UAV-assisted integrated TN-NTN, as shown in Figs. \ref{fig_cap_comparison}–\ref{fig_cap_vs_JI_ratio}. Moreover, with its two algorithms, our proposed framework provides a trade-off analysis between maximizing the total capacity and ensuring fairness among users. In the real world, Algorithm \ref{alg1} enables numerous \textit{smart city} applications with bulk data transfer requirements, where aggregate performance is more important than individual user performance. Similarly, Algorithm \ref{alg2} enables critical \textit{healthcare} applications with guaranteed and equitable QoS requirements for all links. From a future perspective, the proposed framework provides a foundation for the emerging highly dynamic HAP-supported integrated TN-NTN environments, characterized by MC-enabled heterogeneous UAVs with differentiated QoS requirements and dynamic spectrum sharing. Additionally, it triggers further research in this direction, particularly in optimizing the MC links, more complex mobility patterns, dynamic resource sharing, and a robust handover mechanism of UAVs within integrated TN-NTN architectures.

\textcolor{red}{\textit{Scalability Analysis}: 
Both proposed algorithms are scalable in terms of computational complexity as the network size increases. For instance, Algorithm \ref{alg1} evaluates $C^{*}_{i,j}$ for all $JI$ candidate HCU--LCU pairs.
For each pair, the optimal power allocation is obtained through bisection search on a monotone function, which converges in
$\mathcal{O}(JI\log(1/\varepsilon))$ steps for a target accuracy $\varepsilon$.
After computing all $C^{*}_{i,j}$ values, the optimal one-to-one pairing is obtained by applying the Hungarian method, which has a computational complexity of $\mathcal{O}(I^3)$.
Consequently, the overall complexity of Algorithm \ref{alg1} is $\mathcal{O}\!\left(JI\log(1/\varepsilon) + I^3\right),$
which grows polynomially with the network size. \\
Similarly, Algorithm \ref{alg2} further performs a fairness-oriented max--min optimization by sorting the $JI$ candidate values in $\{C^{*}_{i,j}\}$, which incurs a complexity of $\mathcal{O}(JI\log(JI))$.
Subsequently, bisection search is applied over the ordered set $\{C^{*}_{i,j}\}$, where each feasibility check requires solving an assignment problem using the Hungarian method.
Hence, the overall complexity is given by
$
\mathcal{O}\!\left(JI\log(1/\varepsilon) + JI\log(JI) + I^3\log I\right).
$
As evident from their computational complexities, both algorithms exhibit polynomial-time complexity and do not include any combinatorial or exponential growth with the number of UAVs.
}

Despite its advantages, the framework has several limitations. The algorithms exploit only slow-varying channel parameters. While this reduces signaling overhead and ensures the mathematical tractability of high-mobility modeling, it limits the potential advantages of adapting to instantaneous small-scale fading. Similarly, the spectrum sharing of each HCU is restricted to only one LCU. While this simplifies the pairing strategy and interference isolation, it leads to spectral resources under-utilization when the number of LCUs exceeds that of HCUs. Additionally, independent power allocation to UAV-UAV, UAV-RBS, and UAV-HAP links is ignored; simplified UAV trajectory and channel modeling is assumed; and the power constraints of UAVs and MCs with respect to multiple RBSs and satellites are not considered. Future extensions may address these issues.


\section{Conclusion} \label{sec_conc}
In this paper, we have proposed a comprehensive resource sharing and allocation framework tailored for heterogeneous MC-enabled UAVs operating in an integrated TN-NTN environment. Considering the challenges imposed by UAV mobility, link heterogeneity, and diverse QoS requirements, we have developed two algorithms to ensure the efficient and equitable use of spectrum and power resources. The first algorithm prioritizes maximizing the total capacity while maintaining reliability. The second algorithm enhances fairness by maximizing the minimum capacity across all connections. 
The proposed algorithms have been extensively investigated and compared against the state-of-the-art approaches. The results have provided design insights, including the tradeoffs among various parameters, and demonstrated the superior performance of the proposed algorithms compared to competing approaches. Moreover, future work will consider more complex interference scenarios, UAV trajectories, many-to-one spectrum sharing, and more realistic channel conditions. \textcolor{red}{While this work focuses on statistical channel modeling suitable for large-scale and fast-mobility UAV networks, extending the framework to explicitly incorporate delay–Doppler domain modeling using orthogonal time-frequency-space or parametric multipath estimation is an interesting direction for future research.
}

\section*{Appendix A}
Let $\mu_{i,j}$ be a reuse pattern chosen arbitrarily, e.g., $\mu_{i,j}=1$, inserting the value of $\gamma_{j,j}^\ell$ into the left-hand side (LHS) of \eqref{eq_opt_1c}, we obtain
\begin{align}
\operatorname{Pr}&\left\{\gamma_{j,j}^\ell \leq \gamma_0^\ell\right\}
= \operatorname{Pr}\left\{ \frac{P_j^\ell ~ \alpha_{j,j} ~ g_{j,j}^{(s)}}{P_N + P_i^h ~ \alpha_{i, j} ~ g_{i, j}^{(s)}} \leq \gamma_0^\ell \right\} \nonumber\\
&= \operatorname{Pr}\left\{ g_{j,j}^{(s)} \leq \frac{\gamma_0^\ell \left( P_N + P_i^h ~ \alpha_{i, j} ~ g_{i, j}^{(s)} \right)}{P_j^\ell ~ \alpha_{j,j}} \right\}.
\end{align}
Assuming $g_{j,j}^{(s)}$ and $g_{i,j}^{(s)}$ to be independent and identically distributed (i.i.d.) exponential random variables with unit mean, the probability can be written as
\begin{align}
& \int_{0}^{\infty} \mathrm{d}g_{i, j}^{(s)} \int_{0}^{\frac{\gamma_0^\ell \left( P_N ~+~ P_i^h ~ \alpha_{i, j} ~ g_{i, j}^{(s)} \right)}{P_j^\ell ~\alpha_{j,j}}} e^{-(g_{j,j}^{(s)} + g_{i, j}^{(s)})} \mathrm{d}g_{j,j}^{(s)} \nonumber\\
&= \int_{0}^{\infty} \left[ 1 - e^{- \frac{\gamma_0^\ell \left( P_N ~+~ P_i^h~ \alpha_{i, j}~ g_{i, j}^{(s)} \right)}{P_j^\ell ~\alpha_{j,j}} } \right] e^{-g_{i, j}^{(s)}} \mathrm{d}g_{i, j}^{(s)} \nonumber\\
&= 1 - e^{- \frac{\gamma_0^\ell ~ P_N}{P_j^\ell~ \alpha_{j,j}} } \int_{0}^{\infty} e^{-g_{i, j}^{(s)} \left( 1 + \frac{\gamma_0^\ell ~ P_i^h ~\alpha_{i, j}}{P_j^\ell ~ \alpha_{j,j}} \right) } \mathrm{d}g_{i, j}^{(s)} \nonumber\\
&= 1 - e^{- \frac{\gamma_0^\ell ~ P_N}{P_j^\ell ~ \alpha_{j,j}} } \left( \frac{1}{1 + \frac{\gamma_0^\ell ~ P_i^h ~ \alpha_{i, j}}{P_j^\ell ~ \alpha_{j,j}} } \right) \nonumber\\
&= 1 - \frac{ P_j^\ell ~ \alpha_{j,j} ~ e^{- \frac{\gamma_0^\ell ~ P_N}{P_j^\ell ~ \alpha_{j,j}} } }{ P_j^\ell ~ \alpha_{j,j} + \gamma_0^\ell ~ P_i^h ~ \alpha_{i, j} }.
\end{align}
Moreover, the reliability constraint requires that this probability is less than or equal to $P_o$, which is expressed as
\begin{align}
1 - \frac{ P_j^\ell ~ \alpha_{j,j} ~ e^{- \frac{\gamma_0^\ell ~ P_N}{P_j^\ell ~ \alpha_{j,j}} } }{ P_j^\ell ~ \alpha_{j,j} + \gamma_0^\ell ~ P_i^h ~ \alpha_{i, j} } \leq P_o.
\end{align}
Rearranging the terms yields
\begin{align}
P_i^h \leq \frac{ \alpha_{j,j} ~ P_j^\ell }{ \gamma_0^\ell ~ \alpha_{i, j} } \left( \frac{ e^{- \frac{\gamma_0^\ell ~ P_N}{P_j^\ell ~ \alpha_{j,j}} } }{1 - P_o } - 1 \right).
\end{align}
\qed

\section*{Appendix B}
\begin{figure}
\centering
\subfigure[Scenario 1]
{\includegraphics[width=1\linewidth]{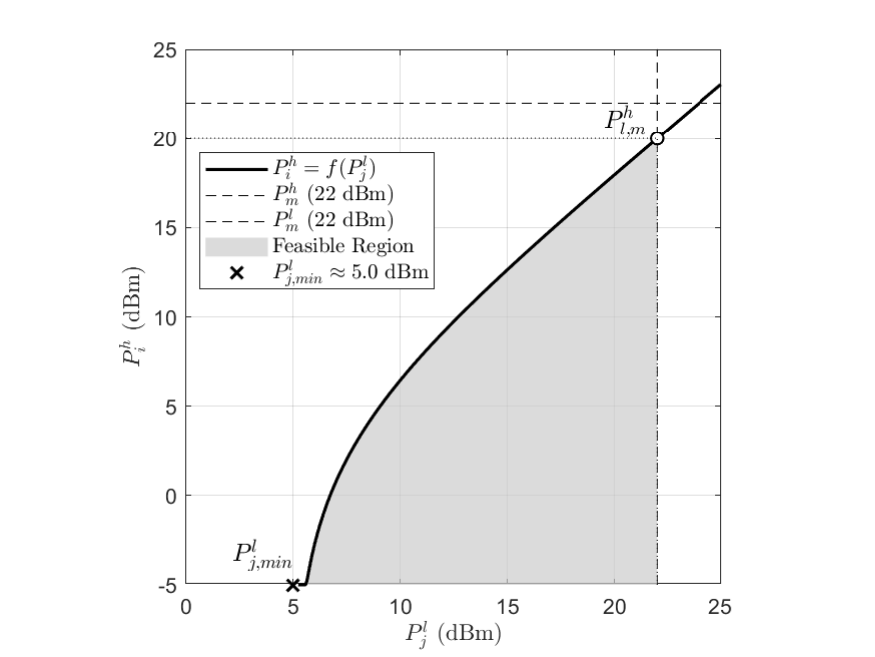}}
\subfigure[Scenario 2]
{\includegraphics[width=1\linewidth]{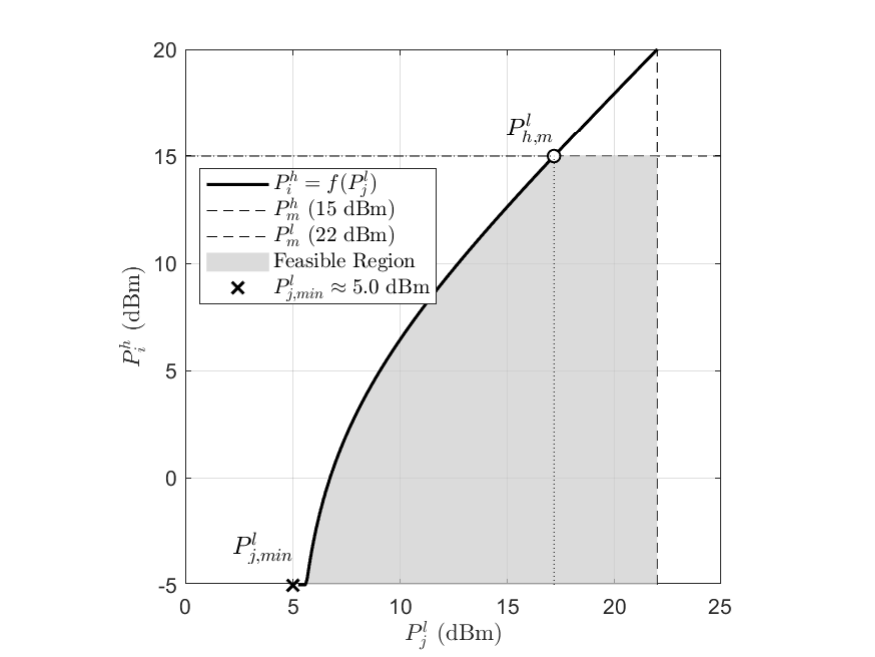}}
\caption{The feasible regions for \eqref{eq_opt_2} plotted as functions of $P^h_m$ and $P^\ell_m$.}
\label{fig_feasible}
\end{figure}

To determine the feasible range of LCU power, we identify the zero-crossing point of the function $f(P_j^\ell)$. This point represents the minimum value of $P_j^\ell$ that satisfies the reliability constraint given in \eqref{eq_opt_2}. We denote this minimum value as $P_{j,\min}^\ell$.  
Since $P_m^h \ge 0$, from \eqref{eq_4_f()}, the zero-crossing point is obtained by nullifying $f(P_j^\ell) = 0$, which gives
\begin{equation}
P_j^\ell = -\frac{\gamma_0^\ell P_N}{\alpha_{j,j} \ln(1-P_o)} \triangleq P_{j,\min}^\ell.
\end{equation}
In addition to \eqref{eq_4_f()}, Fig. \ref{fig_feasible}, which illustrates the feasible regions corresponding to \eqref{eq_opt_2}, also confirms that $f(P_j^\ell)$ is a monotonically increasing function of $P_j^\ell$  over the interval $ [P_{j,\min}^\ell, +\infty) $. This implies that increasing the LCU power widens the interference margin, making the LCU more tolerant to interference from HCUs. The feasible regions in Fig. \ref{fig_feasible} are divided into two scenarios depending on the levels of $P_{m}^h$ and $P_{m}^\ell$.

Now, defining $g_{i,R}^{(s)}$ and $g_{j,R}^{(s)}$, as in Appendix A, the capacity of the $i$th HCU when sharing spectrum with the $j$th LCU can be given by
\begin{align} 
\label{eq_1_app_B}
C_{i,j}^h(P^h_i,P^\ell_j)  & = \mathbb{E}[\log_2(1 + \gamma_{i,R}^h)] \nonumber\\
& =\int_{0}^{\infty}\!\int_{0}^{\infty}
\log_{2}\!\Bigl(1+\frac{P^h_i\,\alpha_{i,R}\,g_{i,R}^{(s)}}{P_N+P^\ell_j\,\alpha_{j,R}\,g_{j,R}^{(s)}}\Bigr) \nonumber\\
&\times e^{-\bigl(g_{i,R}^{(s)}+g_{j,R}^{(s)}\bigr)}\,\mathrm{d}g_{i,R}^{(s)}\,\mathrm{d}g_{j,R}^{(s)}.
\end{align}

From \eqref{eq_1_app_B}, it is observed that for fixed $P^\ell_j$, $C_{i,j}^h(P^h_i,P^\ell_j)$ increases monotonically with $P^h_i$,  and for fixed $P^h_i$, $C_{i,j}^h(P^h_i,P^\ell_j)$ decreases monotonically with $P^\ell_j$.  
Based on these observations, it can be easily inferred that the optimal solution of \eqref{eq_opt_2} is necessarily located at the upper boundary of the shaded region characterized by the curve $ P_i^h = f(P_j^\ell) $, which extends from the point $ (P_{j,\min}^\ell, 0) $ to either $ (P_{m}^\ell, P_{\ell,m}^h) $ in Scenario 1 or $ (P_{h,m}^\ell, P_{m}^h) $ in Scenario 2, as shown in Fig. \ref{fig_feasible}. This inference is supported by the fact that the function $ P_i^h = f(P_j^\ell) $ increases monotonically within the interval $ (P_{j,\min}^\ell, +\infty) $.
Along this boundary, substituting $P^h_i = f(P^\ell_j)$ into \eqref{eq_1_app_B} yields
\begin{align}
\gamma^h_i  &=\frac{P^h_i\,\alpha_{i,R}\,g_{i,R}^{(s)}}{P_N + P^\ell_j\,\alpha_{j,R}\,g_{j,R}^{(s)}}\nonumber\\
&=\frac{\alpha_{j,j}\,\alpha_{i,R}\,g_{i,R}^{(s)}}{\gamma^\ell_0\,\alpha_{i,j}\Bigl(\frac{P_N}{P^\ell_j}+\alpha_{j,R}\,g_{j,R}^{(s)}\Bigr)}
\Biggl(\frac{e^{-\frac{\gamma^\ell_0\,P_N}{P^\ell_j\,\alpha_{j,j}}}}{1-P_o}-1\Biggr).
\end{align}
Therefore, the solution to problem \eqref{eq_opt_2} lies at the intersection point $ (P_{m}^\ell, P_{\ell,m}^h) $ in Scenario 1 or at the intersection point $ (P_{h,m}^\ell, P_{m}^h) $ in Scenario 2. This result is written in short form in \eqref{eq_powers}.
\qed

\section*{Appendix C}
The capacity $ C_{i,j}^h(P^h_i,P^\ell_j) $ can be written as
\begin{equation}
\begin{split}
	C_{i,j}^h(P^h_i,P^\ell_j) & = \mathbb{E} \left[ \log_2 \left( 1 + \frac{P_i^h \alpha_{i,R} g_{i,R}^{(s)}}{P_N + P_j^\ell \alpha_{j,R} g_{j,R}^{(s)}} \right) \right] \\ &
	\triangleq \mathbb{E} \left[ \log_2 \left( 1 + \frac{\rho U}{1 + \eta V} \right) \right],
\end{split}
\end{equation}
The notations $ U $ and $ V $ represent  $ g_{i,R}^{(s)} $ and $ g_{j,R}^{(s)} $, respectively, while 
$\rho$ is defined as
\[
\rho = \frac{P_i^h \alpha_{i,R}}{P_N}, \quad \eta = \frac{P_j^\ell \alpha_{j,R}}{P_N}.
\]
Suppose that $ W = \frac{\rho U}{1 + \eta V} $ and $ g_{i,R}^{(s)} $ and $ g_{j,R}^{(s)} $ are defined as in Appendix A. Then, we can derive its CDF as
\begin{align}
F_W(w) & = \Pr\left\{ \frac{\rho U}{\eta + V1} \leq w \right\} \notag \\
& = \int_{0}^{\infty} \mathrm{d}v \int_{0}^{\frac{w(1 + \eta v)}{\rho}} e^{-(u + v)} \mathrm{d}u \notag \\
& = 1 - e^{-\frac{w}{\rho}} \cdot \frac{\rho}{\rho + \eta w}.
\end{align}
Subsequently, the capacity of the $ i $th HCU is obtained as
\begin{align}
C_{i,j}^h&(P^h_i,P^\ell_j)
= \frac{1}{\ln 2} \int_{0}^{\infty} \ln(1 + w) f_W(w) \, \mathrm{d}w \label{eq_1_app_C} \notag \\
&= \frac{1}{\ln 2} \int_{0}^{\infty} \frac{1 - F_W(w)}{1 + w} \, \mathrm{d}w  \\
&= \frac{\rho}{(\rho - \eta)\ln 2} \left[ \int_0^{\infty} \frac{e^{-w/\rho}}{w + 1} \, \mathrm{d}w - \int_0^{\infty} \frac{e^{-w/\rho}}{w + \frac{\rho}{\eta}} \, \mathrm{d}w \right] \notag \\
&= \frac{\rho}{(\rho - \eta)\ln 2} \left[ e^{\frac{1}{\rho}} E_1\left( \frac{1}{\rho} \right) - e^{\frac{1}{\eta}} E_1\left( \frac{1}{\eta} \right) \right],
\label{eq_2_app_C}
\end{align}
which is obtained through  integration by parts using \cite[eq.~(3.352.4)]{IM:07:B_E}.
\qed

\section*{Acknowledgment}
This work has been supported in part by FWO, Belgium, under file no. V507025N. 

\balance

\bibliographystyle{IEEEtran}
\bibliography{
IEEEabrv, 
./CQILABInclusion/CQILAB-abrv,
./CQILABInclusion/CQILAB-Journal
}


\end{document}